\documentclass[sigconf, nonacm]{acmart}
\usepackage[linesnumbered, ruled, vlined, commentsnumbered, boxed]{algorithm2e}
\usepackage{graphicx}
\usepackage{amsmath}
\usepackage{amsthm}
\usepackage{float}
\usepackage{verbatim}
\usepackage{tipa}
\usepackage{color}
\usepackage{latexsym}
\usepackage{subfig}
\usepackage{multirow}
\usepackage{geometry}
\usepackage{capt-of}
\usepackage{threeparttable}
\usepackage{xcolor}
\newtheorem{lemma}{Lemma}
\newtheorem{theorem}{Theorem}
\newtheorem{definition}{Definition}
\newtheorem{example}{Example}
\newtheorem{property}{Property}
\newcommand\vldbdoi{XX.XX/XXX.XX}

\newcommand\vldbvolume{14}
\newcommand\vldbissue{1}

\begin{document}
	\title{Scalable Time-Range $k$-Core Query on Temporal Graphs}
	
	%%
	%% The "author" command and its associated commands are used to define the authors and their affiliations.
	\author{Junyong Yang}
	\affiliation{%
		\institution{Wuhan University}
		\city{Wuhan}
		\country{China}
	}
	\email{thomasyang@whu.edu.cn}
	\author{Ming Zhong}
	\authornote{The corresponding author.}
	\affiliation{%
		\institution{Wuhan University}
		\city{Wuhan}
		\country{China}
	}
	\email{clock@whu.edu.cn}
	\author{Yuanyuan Zhu}
	\affiliation{%
		\institution{Wuhan University}
		\city{Wuhan}
		\country{China}
	}
	\email{yyzhu@whu.edu.cn}
	\author{Tieyun Qian}
	\affiliation{%
		\institution{Wuhan University}
		\city{Wuhan}
		\country{China}
	}
	\email{qty@whu.edu.cn}
	\author{Mengchi Liu}
	\affiliation{%
		\institution{South China Normal University}
		\city{Guangzhou}
		\country{China}
	}
	\email{liumengchi@scnu.edu.cn}
	\author{Jeffrey Xu Yu}
	\affiliation{%
		\institution{The Chinese University of Hong Kong}
		\city{Hong Kong}
		\country{China}
	}
	\email{yu@se.cuhk.edu.hk}
	
	%%
	%% The abstract is a short summary of the work to be presented in the
	%% article.
	\begin{abstract}
		Querying cohesive subgraphs on temporal graphs with various time constraints has attracted intensive research interests recently. In this paper, we study a novel Temporal $k$-Core Query (TCQ) problem: given a time interval, find all distinct $k$-cores that exist within any subintervals from a temporal graph, which generalizes the previous historical $k$-core query. This problem is challenging because the number of subintervals increases quadratically to the span of time interval. For that, we propose a novel Temporal Core Decomposition (TCD) algorithm that decrementally induces temporal $k$-cores from the previously induced ones and thus reduces ``intra-core'' redundant computation significantly. Then, we introduce an intuitive concept named Tightest Time Interval (TTI) for temporal $k$-core, and design an optimization technique with theoretical guarantee that leverages TTI as a key to predict which subintervals will induce duplicated $k$-cores and prunes the subintervals completely in advance, thereby eliminating ``inter-core'' redundant computation. The complexity of optimized TCD (OTCD) algorithm no longer depends on the span of query time interval but only the scale of final results, which means OTCD algorithm is scalable. Moreover, we propose a compact in-memory data structure named Temporal Edge List (TEL) to implement OTCD algorithm efficiently in physical level with bounded memory requirement. TEL organizes temporal edges in a ``timeline'' and can be updated instantly when new edges arrive, and thus our approach can also deal with dynamic temporal graphs. We compare OTCD algorithm with the incremental historical $k$-core query on several real-world temporal graphs, and observe that OTCD algorithm outperforms it by three orders of magnitude, even though OTCD algorithm needs none precomputed index.
	\end{abstract}
	
	\maketitle
	
	%%% do not modify the following VLDB block %%
	%%% VLDB block start %%%
	\begingroup
	\renewcommand\thefootnote{}\footnote{\noindent
		This work is licensed under the Creative Commons BY-NC-ND 4.0 International License. Visit \url{https://creativecommons.org/licenses/by-nc-nd/4.0/} to view a copy of this license. For any use beyond those covered by this license, obtain permission by emailing \href{mailto:info@vldb.org}{info@vldb.org}. Copyright is held by the owner/author(s). Publication rights licensed to the VLDB Endowment. \\
		\raggedright Proceedings of the VLDB Endowment, Vol. \vldbvolume, No. \vldbissue\ %
		ISSN 2150-8097. \\
		\href{https://doi.org/\vldbdoi}{doi:\vldbdoi} \\
	}\addtocounter{footnote}{-1}\endgroup
	%%% VLDB block end %%%
	
\section{INTRODUCTION}
	
\subsection{Motivation}
	
\begin{figure}[t!]
		\centering
		\includegraphics[width=\linewidth]{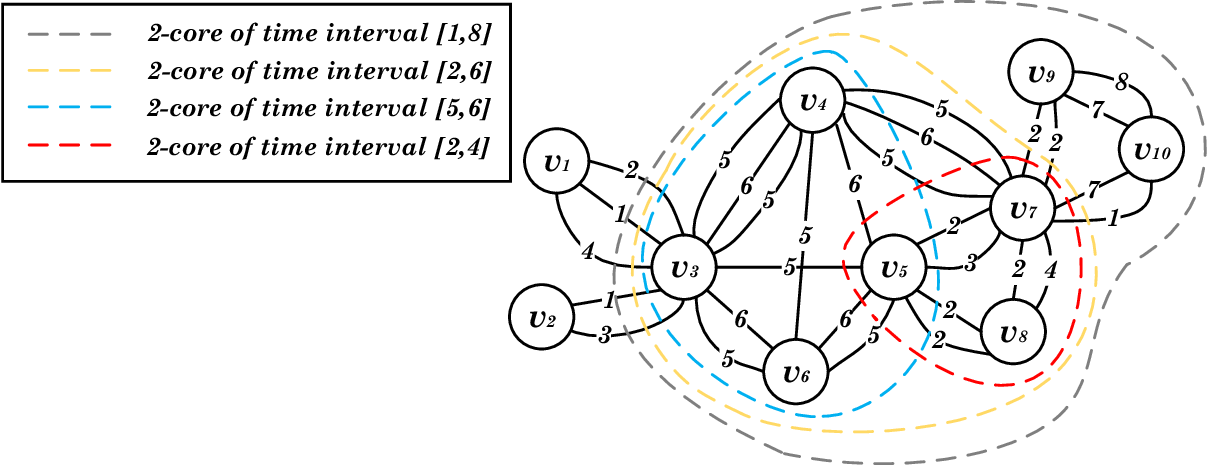}
		\caption{A running example of temporal graph.}\label{fig:temporalgraph}
	\end{figure}

Discovering communities or cohesive subgraphs from temporal graphs has great values in many application scenarios, thereby attracting intensive research interests~\cite{yu2021querying, bai2020efficient, galimberti2018mining, wu2015core, chu2019online, qin2020periodic, li2018persistent, ma2019efficient} in recent years. Here, a temporal graph refers to an undirected multigraph in which each edge has a timestamp to indicate when it occurred, as illustrated in Figure~\ref{fig:temporalgraph}. For example, consider a graph consisting of bank accounts as vertices and fund transfer transactions between accounts as edges with natural timestamps. For applications such as anti-money-laundering, we would like to search communities like $k$-cores that contain a known suspicious account and emerge within a specific time interval like the World Cup, and investigate the associated accounts.
	
To address the community query/search problem for a fixed time interval, the concept of historical $k$-core~\cite{yu2021querying} is proposed recently, which is the $k$-core induced from the subgraph of a temporal graph in which all edges occurred out of the time interval have been excluded and the parallel edges between each pair of vertices have been merged. Also, the PHC-Query method is proposed to deal with historical $k$-core query/search by using a precomputed index efficiently.
	
However, we usually do not know the exact time interval of targeted historical $k$-core in real-world applications. Actually, if we can know the exact time interval, a traditional core decomposition on the projected graph over the given time interval is efficient enough to address the problem. Thus, it is more reasonable to assume that we can only offer a flexible time interval and need to induce cores from all its subintervals. For example, for detecting money laundering by soccer gambling during the World Cup, the $k$-cores emerged over a few of hours around one of the matches are more valuable than a large $k$-core emerging over the whole month. 

Therefore, we aim to generalize historical $k$-core query by allowing the result $k$-cores to be induced by any subinterval of a given time interval, like ``flexible versus fixed''. The historical $k$-core query can be seen as a special case of our problem that only evaluates the whole interval. Consider the following example.

\begin{example}
As illustrated in Figure~\ref{fig:temporalgraph}, given a time interval [1,8], historical $k$-core query only returns the largest core marked by the grey dashed line. In contrast, our temporal $k$-core query returns four cores marked by dashed lines with different colors. These cores can reveal various insights unseen by the largest one. For example, some cores like red and blue that emerge in bursty periods may be caused by special events. Also, some persistent or periodic cores may be found. Further, we can analyze the interaction between cores and how they evolve over time, such as the small cores like red and blue are merged to the large cores like yellow. Lastly, some underlying details may be found. During the merge, the vertex $v_5$ may play a vital role because it appears in all  the cores. Overall, our general and flexible query model can support many interesting temporal community analytics applications.
\end{example}

The general and flexible temporal k-core query we study is naturally a generalization of existing query models like historical $k$-core and also potentially a common technique for various temporal graph mining tasks mentioned in the above example.

\subsection{Contribution}

In this paper, we study a novel \textit{temporal $k$-core query} problem: given a time interval, find all distinct $k$-cores that exist within any subintervals from a temporal graph. Although the existing PHC-Query returns the historical $k$-core of a fixed time interval efficiently, it cannot be trivially applied to deal with the new problem. Because inducing $k$-cores for each subinterval individually from scratch is not scalable, since the number of subintervals increases quadratically with the span of time interval. Moreover, PHC-Query suffers from two other intrinsic shortcomings. Firstly, it relies on a PHC-Index that precomputes the coreness of all vertices over all time intervals, thereby incurring heavy offline time and space overheads. Secondly, due to its sophisticated construction, it is unclear if PHC-Index can be updated dynamically. It is against the dynamic nature of temporal graphs.
	
In order to overcome the above challenges, we present a novel \textit{temporal core decomposition} algorithm and auxiliary optimization and implementation techniques. Our contributions can be summarized as follows.

	\begin{itemize}
		\item We formalize a general time-range cohesive subgraph query problem on ubiquitous temporal graphs, namely, temporal $k$-core query. Many previous typical $k$-core query models on temporal graphs can be equivalently represented by temporal $k$-core query with particular constraints.
		\item To address temporal $k$-core query, we propose a simple and yet efficient algorithm framework based on a novel temporal core decomposition operation. By using temporal core decomposition, our algorithm always decrementally induces a temporal k-core from the previous induced temporal k-core except the initial one, thereby reducing redundant computation significantly.
		\item Moreover, we propose an intuitive concept named tightest time interval for temporal k-core. According to the properties of tightest time intervals, we design three pruning rules with theoretical guarantee to directly skip subintervals that will not induce distinct temporal $k$-core. As a result, the optimized algorithm is scalable in terms of the span of query time interval, since only the necessary subintervals are enumerated.
		\item For physical implementation of our algorithm, we propose a both space and time efficient data structure named temporal edge list to represent a temporal graph in memory. It can be manipulated to perform temporal core decomposition and tightest time interval based pruning rapidly with bounded memory. More importantly, temporal edge list can be incrementally updated with evolving temporal graphs, so that our approach can support dynamic graph applications naturally.
		\item Lastly, we evaluate the efficiency and effectiveness of our algorithm on real-world datasets. The experimental results demonstrate that our algorithm outperforms the improved PHC-Query by three orders of magnitude.
	\end{itemize}
	
The rest of this paper is organized as follows. Section 2 formally introduces the data model and query model, and also gives a baseline algorithm. Sections 3-5 present our algorithm, optimization and implementation techniques respectively. Section 6 briefly discusses some meaningful extension of our approach. Section 7 presents the experiments and analyzes the results. Section 8 investigates the related work. Section 9 concludes our work.
	
\section{Preliminary}\label{sec:problem}
	
In this section, we propose a generalized $k$-core query problem on temporal graphs, which facilitates various temporal community query/search demands. The previous historical $k$-core query~\cite{yu2021querying} can be seen as a special case of the proposed problem. Specifically, we introduce the data model and query model of the proposed problem in Section~\ref{sec:datamodel} and~\ref{sec:querymodel} respectively, and then present a nontrivial baseline that addresses the proposed problem based on the existing PHC-Query.
	
\subsection{Data Model}\label{sec:datamodel}
	
A \textit{temporal graph} is normally an undirected graph $\mathcal{G} = (\mathcal{V}, \mathcal{E})$ with parallel temporal edges. Each temporal edge $(u, v, t) \in \mathcal{E}$ is associated with a timestamp $t$ that indicates when the interaction happened between the vertices $u, v \in \mathcal{V}$. For example, the temporal edges could be transfer transactions between bank accounts in a finance graph. Without a loss of generality, we use continuous integers that start from 1 to denote timestamps. Figure~\ref{fig:temporalgraph} illustrates a temporal graph as our running example.
	
There are two useful concepts derived from the temporal graph. Given a time interval $[ts, te]$, we define the \textit{projected graph} of $\mathcal{G}$ over $[ts, te]$ as $\mathcal{G}_{[ts, te]} = (\mathcal{V}_{[ts, te]}, \mathcal{E}_{[ts, te]})$, where $\mathcal{V}_{[ts, te]} = \mathcal{V}$ and $\mathcal{E}_{[ts, te]} = \{(u, v, t) | (u, v, t)\in \mathcal{E}, t\in[ts, te]\}$. Moreover, we define the \textit{detemporalized graph} of $\mathcal{G}_{[ts, te]}$ as a simple graph $G_{[ts, te]} = (V_{[ts, te]}, E_{[ts, te]})$, where $V_{[ts, te]}$=$\mathcal{V}_{[ts, te]}$ and $E_{[ts, te]}$ = $\{(u, v) | (u, v, t) \in \mathcal{E}_{[ts, te]}\}$.
	
\subsection{Query Model}\label{sec:querymodel}
	
For revealing communities in graphs, the $k$-core query is widely adopted. Given an undirected graph $G$ and an integer $k$, $k$-core is the maximal induced subgraph of $G$ in which all vertices have degrees at least $k$, which is denoted by $\mathcal{C}^{k}(G)$. The \textit{coreness} of a vertex $v$ in a graph $G$ is the largest value of $k$ such that $v \in \mathcal{C}^{k}(G)$.
	
For temporal graphs, the Historical $k$-Core Query (HCQ)~\cite{yu2021querying} is proposed recently. It aims to find a $k$-core that appears during a specific time interval. Formally, a historical $k$-core $\mathcal{H}_{[ts, te]}^{k}(\mathcal{G})$ is a $k$-core in the detemporalized projected graph $G_{[ts, te]}$ of $\mathcal{G}$. Thus, HCQ can be defined as follows.
	
	\begin{definition}[Historical $k$-Core Query]
		For a temporal graph $\mathcal{G}$, given an integer $k$ and a time interval $[ts, te]$, return $\mathcal{H}_{[ts, te]}^{k}(\mathcal{G}) = \mathcal{C}^{k}(G_{[ts,te]})$.
	\end{definition}
	
In this paper, we propose a novel query model called Temporal $k$-Core Query (TCQ) that generalizes HCQ. The main difference is that the query time interval $[Ts, Te]$ of TCQ is a range but not fixed query condition like $[ts,te]$ of HCQ. In TCQ, $Ts$ and $Te$ are the minimum start time and maximum end time of query time interval respectively, and thereby the $k$-cores induced by each subinterval $[ts, te] \subseteq [Ts, Te]$ are all potential results of TCQ. Moreover, TCQ directly returns the maximal induced subgraphs of $\mathcal{G}$ in which all vertices have degrees (note that, the number of neighbor vertices but not neighbor edges) at least $k$ as results. We call these subgraphs as \textit{temporal $k$-cores} and denote by $\mathcal{T}^k_{[ts,te]}(\mathcal{G})$ a temporal $k$-core that appears over $[ts,te]$ on $\mathcal{G}$. Obviously, a historical $k$-core $\mathcal{H}_{[ts, te]}^{k}(\mathcal{G})$ is the detemporalized temporal $k$-core $\mathcal{T}^k_{[ts,te]}(\mathcal{G})$. Therefore, TCQ can be seen as a group of HCQ and HCQ can be seen as a special case of TCQ. 
	
The formal definition of TCQ is as follows.

	\begin{definition}[Temporal $k$-Core Query]\label{def:tcq}
		For a temporal graph $\mathcal{G}$, given an integer $k$ and a time interval $[Ts, Te]$, return all distinct $\mathcal{T}_{[ts, te]}^{k}(\mathcal{G})$ with $[ts, te] \subseteq [Ts, Te]$.
	\end{definition}
	
Note that, TCQ only returns the distinct temporal $k$-cores that are not identical to each other, since multiple subintervals of $[Ts,Te]$ may induce an identical subgraph of $\mathcal{G}$. For brevity, $\mathcal{T}_{[ts,te]}^{k}(\mathcal{G})$ is abbreviated as $\mathcal{T}_{[ts,te]}^{k}$ if the context is self-evident.

\subsection{Baseline Algorithm}\label{sec:baseline}
	
A straightforward solution to TCQ is to enumerate each subinterval $[ts, te] \subseteq [Ts, Te]$ and induce $\mathcal{T}_{[ts, te]}^{k}$ respectively, which takes $O(|Te-Ts|^{2}|\mathcal{E}|)$ time. However, the span of query time interval (namely, $Te-Ts$) can be extremely large in practice, which results in enormous time consumption for inducing all temporal $k$-cores from scratch independently. Therefore, we start from a non-trivial baseline based on the existing PHC-Query.
	
\subsubsection{A Short Review of PHC-Query}
	
PHC-Query relies on a heavyweight index called PHC-Index that essentially precomputes the coreness of all vertices in the projected graphs over all possible time intervals. The index is logically a table that stores a set of timestamp pairs for each vertex $v \in \mathcal{V}$ (column) and each reasonable coreness $k$ (row). Given a value of $k$, the coreness of a vertex $v$ is exactly $k$ in the projected graph over $[ts, te]$ for each timestamp pair $ts$ and $te$ in the cell ($k,v$). In particular, due to the monotonicity of coreness of a vertex with respect to $te$ when $ts$ is fixed, PHC-Index can reduce its space cost significantly by only storing the necessary but not all possible timestamp pairs. Specifically, for a vertex $v$, a coreness $k$ and a start time $ts$, only a discrete set of \textit{core time} need to be recorded, since the coreness of the vertex over $[ts, te]$ will not change with the increase of $te$ until $te$ is a core time. Consequently, given an HCQ instance, PHC-Query leverages PHC-Index to directly determine whether a vertex has the coreness no less than the required $k$, by comparing the query time interval with the retrieved timestamp pairs, and then induces historical $k$-cores with qualified vertices.
	
\subsubsection{Incremental PHC-Query Algorithm}
	
The main idea of our baseline algorithm is to induce temporal $k$-cores incrementally, thereby reducing redundant computation. With a temporal $k$-core $\mathcal{T}^k_{[ts,te]}$, we induce $\mathcal{T}^k_{[ts,te+1]}$ simply by appending new vertices to $\mathcal{T}^k_{[ts,te]}$, whose coreness has become no less than $k$ due to the expand of time interval. Those vertices can be directly identified by using core time retrieved from PHC-Index since $ts$ is fixed. The correctness of baseline algorithm is guaranteed while the correctness of PHC-Query holds.

The pseudo code of incremental PHC-Query (iPHC-Query) algorithm is presented in Algorithm~\ref{alg:baseline}. It enumerates all subintervals of a given $[Ts, Te]$ in a particular order for fulfilling efficient incremental temporal $k$-core induction. Specifically, it anchors the value of $ts$ (line 1), and increases the value of $te$ from $ts$ to $Te$ (line 5), so that $\mathcal{T}^k_{[ts,te+1]}$ can always be incrementally generated from an existing $\mathcal{T}^k_{[ts,te]}$. For each $ts$ anchored and the input $k$, the algorithm firstly retrieves the core time of all vertices from PHC-Index, and pushes the vertices into a minimum heap $\mathbb{H}_v$ ordered by their core time (line 3). Moreover, all temporal edges with timestamps in $[ts, Te]$ are pushed into another minimum heap $\mathbb{H}_e$ ordered by their timestamp (line 4). Then, the algorithm maintains a vertex set $\mathbb{V}$ and an edge set $\mathbb{E}$, which represent the vertices and edges of $\mathcal{T}_{[ts,te]}^{k}$ respectively, whenever $te$ is increased by the following steps. It pops remaining vertices with core time no greater than $te$ from $\mathbb{H}_v$ and adds them to $\mathbb{V}$ (line 6), since the corenesss of these vertices are no less than $k$ according to PHC-Index. Also, it pops remaining edges with timestamp no greater than $te$ from $\mathbb{H}_e$ and adds them to $\mathbb{E}$ if both vertices linked by the edges are in $\mathbb{V}$ (line 7). Then, it puts back the popped edges that are not in $\mathbb{E}$ into $\mathbb{H}_e$ (line 8), because they could still be contained by other temporal $k$-cores induced later. Lastly, a temporal $k$-core comprised of $\mathbb{V}$ and $\mathbb{E}$ that are not empty is collected if it has not been induced before (line 9).
	
	\begin{algorithm}[t!]
		\DontPrintSemicolon
		\KwIn{$\mathcal{G},\ k,\ Ts,\ Te$ \\
		}
		\KwOut{all distinct $\mathcal{T}_{[ts,te]}^{k}(\mathcal{G})$ with $[ts,te] \subseteq [Ts,Te]$}
		\For{$ts\leftarrow$ $Ts$ \KwTo $Te$}{
			$\mathbb{V} \leftarrow \emptyset$, $\mathbb{E} \leftarrow \emptyset$, $\mathbb{H}_v \leftarrow \emptyset$, $\mathbb{H}_e \leftarrow \emptyset$\;
			for $k$ and $ts$, retrieve the core time of each vertex in $\mathcal{G}$ from PHC-Index and push them into $\mathbb{H}_v$\;
			push the temporal edges with timestamps in $[ts, Te]$ in $\mathcal{G}$ into $\mathbb{H}_e$\;
			\For{$te\leftarrow ts$ \KwTo $Te$}{
				pop a vertex from $\mathbb{H}_v$ and add it to $\mathbb{V}$, until the min core time of $\mathbb{H}_v$ exceeds $te$\;
				pop an edge from $\mathbb{H}_e$ and add it to $\mathbb{E}$ if both vertices linked by this edge are in $\mathbb{V}$, until the min timestamp of $\mathbb{H}_e$ exceeds $te$\;
				push all edges that have been popped from $\mathbb{H}_e$ and are not added to $\mathbb{E}$ back to $\mathbb{H}_e$\;
				collect $\mathcal{T}_{[ts,te]}^{k}$ = $(\mathbb{V}, \mathbb{E})$ if it is neither empty nor identical to other existing results\;
			}
		}
		\caption{Baseline iPHC-Query algorithm.}\label{alg:baseline}
	\end{algorithm}

The complexity of baseline mainly depends on the maintenance of both $\mathbb{V}$ and $\mathbb{E}$. For the maintenance of $\mathbb{V}$, each vertex in $\mathcal{T}_{[ts,Te]}^{k}$is added to $\mathbb{V}$ from $\mathbb{H}_{v}$ at most once in the inner loop (lines 5-9), which takes logarithmic time for a heap. Therefore, the total cost is bounded by $\sum_{t=Ts}^{Te}|\mathcal{V}_{[t,Te]}|\log|\mathcal{V}_{[t,Te]}|$. The case is more complicated for the maintenance of $\mathbb{E}$, since each edge with a timestamp within $[t, Te]$ is likely to be transferred between $\mathbb{H}_{e}$ and $\mathbb{E}$ (lines 7-8), until both its endpoints are contained by $\mathbb{V}$. In the worst case, the total cost is bounded by $\sum_{t=Ts}^{Te}|Te-t||\mathcal{E}_{[t,Te]}|\log|\mathcal{E}_{[t,Te]}|$. While, the real cost in practice can be much lower since the $|Te-t|$ part should be a more reasonable value.

Although the baseline algorithm can achieve incremental induction of temporal k-core for each start time, PHC-Index incurs a huge amount of extra space and time overheads. Moreover, its incremental induction only offers a kind of ``intra-core'' optimization that reduces the redundant computation in each temporal $k$-core induction, and lacks of a kind of ``inter-core'' optimization that can directly avoids inducing some temporal $k$-cores. In the following sections, we first propose a novel algorithm that can outperform baseline algorithm without any precomputation and index, and then optimize it significantly to further improve the efficiency by at least three orders of magnitude.

\section{Algorithm}\label{sec:alg}
	
In this section, we propose a novel efficient algorithm to address TCQ. Our algorithm leverages a fundamental operation called \textit{temporal core decomposition} to induce $\mathcal{T}^k_{[ts,te]}$ from $\mathcal{T}^k_{[ts,te+1]}$ decrementally. More importantly, our algorithm does not require any precomputation and index space, and can still outperform the baseline algorithm. Next, Section~\ref{sec:tcd} introduces the temporal core decomposition operation, and Section~\ref{sec:tcd-basic} presents our algorithm.
	
\subsection{Temporal Core Decomposition (TCD)}\label{sec:tcd}
	
Firstly, we introduce Temporal Core Decomposition (TCD) as a basic operation on temporal graphs, which is derived from the traditional \textit{core decomposition}~\cite{batagelj2003m} on ordinary graphs. TCD refers to a two-step operation of inducing a temporal $k$-core $\mathcal{T}^k_{[ts,te]}$ of a given time interval $[ts,te]$ from a given temporal graph $\mathcal{G}$. The first step is \textit{truncation}: remove temporal edges with timestamps not in $[ts, te]$ from $\mathcal{G}$, namely, induce the projected graph $\mathcal{G}_{[ts,te]}$. The second step is \textit{decomposition}: iteratively peel vertices with degree (the number of neighbor vertices but not neighbor edges) less than $k$ and the edges linked to them together. The correctness of TCD is as intuitive as core decomposition.

An excellent property of TCD operation is that, it can induce a temporal $k$-core $\mathcal{T}^k_{[ts,te]}$ from another temporal $k$-core $\mathcal{T}^k_{[ts',te']}$ with $[ts,te] \subset [ts',te']$, so that we can develop a decremental algorithm based on TCD operation to achieve efficient processing of TCQ. To prove the correctness of this property, let us consider the following Theorem~\ref{thm:tcd}.
	
	\begin{lemma}\label{thm:subgraph}
		Given time intervals $[ts, te]$ and $[ts',te']$ such that $[ts, te]\subset [ts',te']$, we have $\mathcal{T}_{[ts, te]}^{k}$ is a subgraph of $\mathcal{T}_{[ts',te']}^{k}$.
	\end{lemma}
	
	\begin{proof}
		For each vertex in $\mathcal{T}_{[ts,te]}^{k}$, its coreness in $\mathcal{G}_{[ts',te']}$ is certainly no less than in $\mathcal{G}_{[ts,te]}$ (namely, $\geqslant k$), because $\mathcal{G}_{[ts,te]}$ is a subgraph of $\mathcal{G}_{[ts',te']}$. Thus, all vertices in $\mathcal{T}_{[ts,te]}^{k}$ will be contained by $\mathcal{T}_{[ts',te']}^{k}$ that is a temporal $k$-core of $\mathcal{G}_{[ts',te']}$.
	\end{proof}
	
	\begin{theorem}\label{thm:tcd}
		Given a time interval $[ts, te]$ and a temporal $k$-core $\mathcal{T}^k_{[ts',te']}$ with $[ts, te]\subset [ts',te']$, the subgraph induced by using TCD operation from $\mathcal{T}_{[ts', te']}^{k}$ for $[ts,te]$ is $\mathcal{T}_{[ts,te]}^{k}$.
	\end{theorem}
	
	\begin{proof}
		Firstly, we prove for any temporal graph $\mathcal{G}'$ satisfying that $\mathcal{T}_{[ts,te]}^{k}$ is a subgraph of $\mathcal{G}'$ and $\mathcal{G}'$ is a subgraph of $\mathcal{G}$, we can induce $\mathcal{T}_{[ts,te]}^{k}$ from $\mathcal{G}'$ by using TCD operation. For each vertex in $\mathcal{T}_{[ts,te]}^{k}$, its coreness is not less than $k$ in $\mathcal{G}'$ over $[ts,te]$, because this temporal $k$-core is a subgraph of $\mathcal{G}'$. Meanwhile, for each vertex in $\mathcal{G}'$ but not in $\mathcal{T}_{[ts,te]}^{k}$, its coreness in $\mathcal{G}'$ is not greater than in $\mathcal{G}$, because $\mathcal{G}'$ is a subgraph of $\mathcal{G}$. Thus, its coreness in $\mathcal{G}'$ over $[ts,te]$ is less than $k$, because it is not in the temporal $k$-core $\mathcal{T}_{[ts,te]}^{k}$ of $\mathcal{G}$. As a result, $\mathcal{T}_{[ts,te]}^{k}$ is also a temporal $k$-core of $\mathcal{G}'$, and thereby can be induced by using TCD operation from $\mathcal{G}'$.\\
		Then, consider two temporal $k$-cores $\mathcal{T}_{[ts,te]}^{k}$ and $\mathcal{T}_{[ts',te']}^{k}$ with $[ts,te] \subseteq [ts',te']$. Due to Lemma~\ref{thm:subgraph}, we have  $\mathcal{T}_{[ts,te]}^{k}$ is a subgraph of $\mathcal{T}_{[ts',te']}^{k}$. Let $\mathcal{G}'_{[ts,te]}$ be the temporal graph induced by the first step of TCD from $\mathcal{T}_{[ts',te']}^{k}$, which is certainly a subgraph of $\mathcal{T}_{[ts',te']}^{k}$. Since $\mathcal{G}'_{[ts,te]}$ only removes the temporal edges not in $[ts,te]$, which means these edges are not contained by $\mathcal{T}_{[ts,te]}^{k}$, it is obviously $\mathcal{T}_{[ts,te]}^{k}$ is a subgraph of $\mathcal{G}'_{[ts,te]}$. Thus, the correctness of this theorem holds.
	\end{proof}
	
\begin{figure}[t!]
		\centering
		\includegraphics[width=\linewidth]{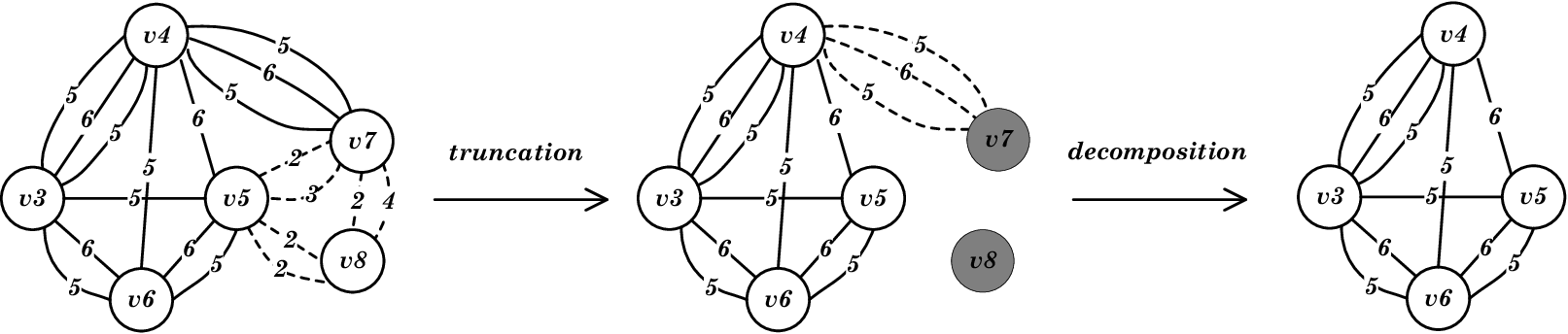}
		\caption{Temporal core decomposition from $\mathcal{T}_{[2,6]}^{2}$ to $\mathcal{T}_{[5,6]}^{2}$.}\label{fig:tdcdstep}
	\end{figure}

For example, Figure~\ref{fig:tdcdstep} illustrates the procedure of TCD from $\mathcal{T}^{2}_{[2,6]}$ to $\mathcal{T}^{2}_{[5,6]}$ on our running example graph in Figure~\ref{fig:temporalgraph}. The edges with timestamps not in $[5,6]$ (marked by dashed lines) are firstly removed from $\mathcal{T}_{[2,6]}^{2}$ by truncation, which results in the decrease of degrees of vertices $v_5$, $v_7$ and $v_8$. Then, the vertices with degree less than 2 (marked by dark circles), namely, $v_7$ and $v_8$ are further peeled by decomposition, together with their edges. The remaining temporal graph is $\mathcal{T}_{[5,6]}^{2}$.
	
	\subsection{TCD Algorithm}\label{sec:tcd-basic}
	
We propose a TCD algorithm to address TCQ by using temporal core decomposition. In general, given a TCQ instance, the TCD algorithm enumerates each subinterval of $[Ts,Te]$ in a particular order, so that the temporal $k$-cores of each subinterval are induced decrementally from previously induced temporal $k$-cores except the initial one.
	
Specifically, we enumerate a subinterval $[ts,te]$ of $[Ts,Te]$ as follows. Initially, let $ts = Ts$ and $te = Te$. It means we induce the largest temporal $k$-core $\mathcal{T}^k_{[Ts,Te]}$ at the beginning. Then, we will anchor the start time $ts = Ts$ and decrease the end time $te$ from $Te$ until $ts$ gradually. As a result, we can always leverage TCD to induce the temporal $k$-core of current subinterval $[ts,te]$ from the previously induced temporal $k$-core of $[ts,te+1]$ but not from $\mathcal{G}_{[ts,te]}$ or even $\mathcal{G}$. Whenever the value of $te$ is decreased to $ts$, the value of $ts$ will be increased to $ts+1$ until $ts = Te$, and the value of $te$ will be reset to $Te$. Then, we induce $\mathcal{T}_{[ts+1,te]}^k$ from $\mathcal{T}_{[ts,te]}^k$, and start over the decremental TCD procedure. The pseudo code of TCD algorithm is given in Algorithm~\ref{alg:tcd}. Note that, the details of TCD($\mathcal{G}$, $k$, $[ts,te]$) function is left to Section~\ref{sec:tcdimpl}, in which we design a specific data structure to implement TCD operation efficiently in physical level.
	
	\begin{algorithm}[t!]
		\DontPrintSemicolon
		\KwIn{$\mathcal{G}$, $k$, $[Ts,Te]$}
		\KwOut{all distinct $\mathcal{T}_{[ts,te]}^{k}$ with $[ts,te] \subseteq [Ts,Te]$}
		\For(\tcp*[f]{anchor a new start time}){$ts\leftarrow$ $Ts$ \KwTo $Te$}{
			$te \leftarrow Te$\tcp*[f]{reset the end time}\;
			\eIf{$ts = Ts$}{
				$\mathcal{T}^k_{[ts,te]}$ $\leftarrow$ TCD($\mathcal{G}_{[Ts,Te]}$, $k$, $[ts,te]$)\;
			}{
				$\mathcal{T}^k_{[ts,te]}$ $\leftarrow$ TCD($\mathcal{T}_{[ts-1,te]}^{k}$, $k$, $[ts,te]$)\;
			}
			collect $\mathcal{T}^k_{[ts,te]}$ if it is distinct\;
			\For(\tcp*[f]{iteratively decremental induction}){$te\leftarrow$ $Te-1$ \KwTo $ts$}{
				$\mathcal{T}_{[ts,te]}^{k}$ $\leftarrow$ TCD($\mathcal{T}_{[ts,te+1]}^{k}$, $k$, $[ts,te]$)\;
				collect $\mathcal{T}^k_{[ts,te]}$ if it is distinct\;
			}
		}
		\caption{TCD algorithm.}\label{alg:tcd}
	\end{algorithm}
	
\begin{figure}[t!]
		\centering
		\includegraphics[width=\linewidth]{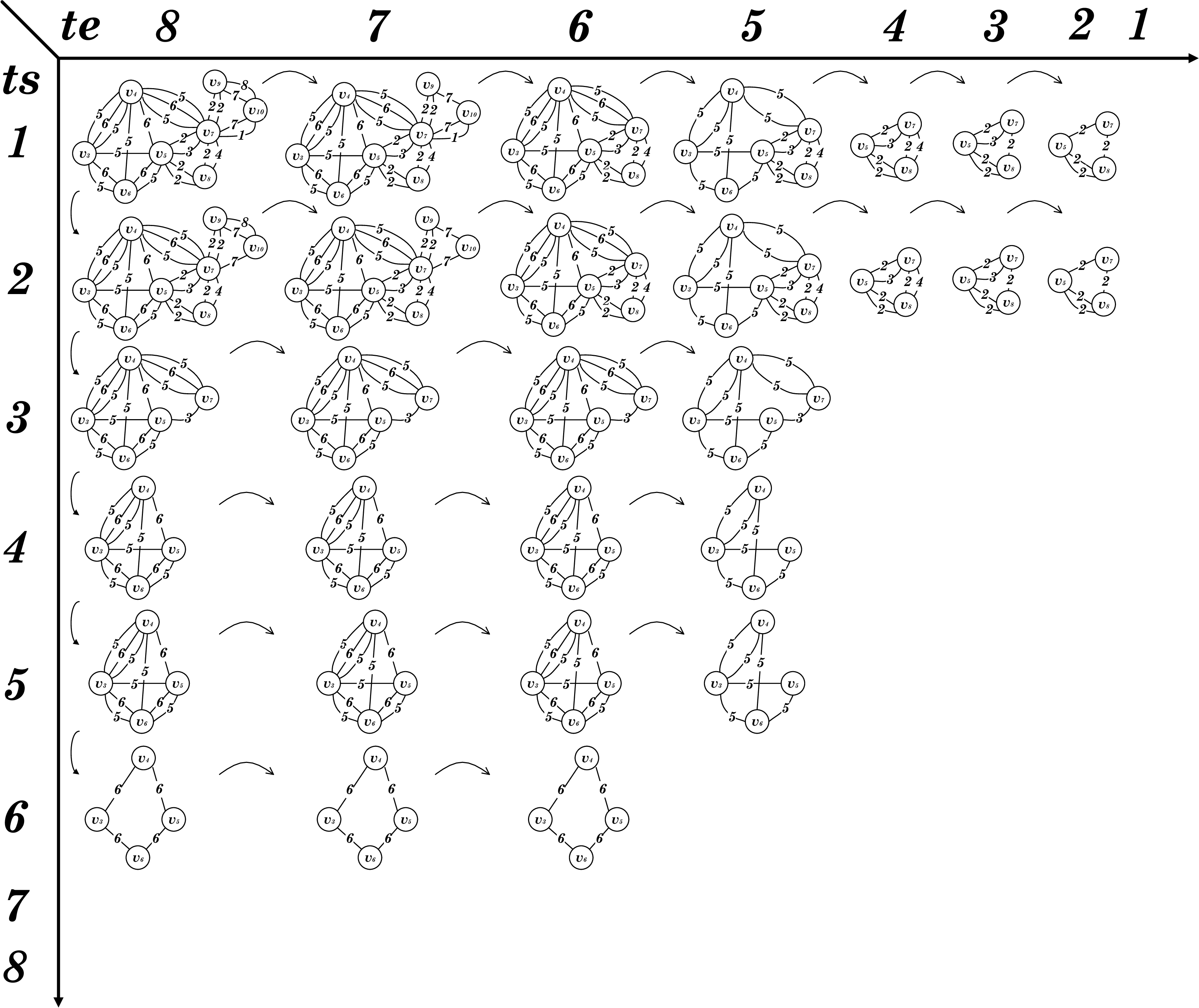}
		\caption{A demonstration of TCD algorithm for finding temporal 2-cores of time interval [1,8].}\label{fig:tcdprocedure}
	\end{figure}

Figure~\ref{fig:tcdprocedure} gives a demonstration of TCD algorithm for finding temporal 2-cores of time interval [1,8] on our running example graph. The temporal $k$-cores are induced line by line and from left to right. Each arrow between temporal $k$-cores represents a TCD operation from tail to head. We can see that, compared with inducing each temporal $k$-core independently, the TCD algorithm reduces the computational overhead significantly. For most induced temporal $k$-cores, a number of vertices and edges have already been excluded while inducing the previous temporal $k$-cores. Moreover, with the increase of $ts$ and the decrease of $te$ when $ts$ is fixed, the size of $\mathcal{T}_{[ts,te]}^k$ will be reduced monotonically until no temporal $k$-core exists over $[ts,te]$, so that the time and space costs of TCD operation will also be reduced gradually.
	
Lastly, we compare TCD algorithm with Baseline algorithm abstractly. When $ts$ is fixed, Baseline algorithm conducts an incremental procedure, in which each vertex is popped once and each edge may be popped and pushed back many times, and in contrast, TCD algorithm conducts a decremental procedure, in which each vertex is peeled once and each edge is also removed once due to Lemma~\ref{thm:subgraph}. Therefore, TCD algorithm that is well implemented in physical level (see Section~\ref{sec:tcdimpl}) can be even more efficient than Baseline algorithm, though it does not need any precomputed index.

\section{Optimization}
	
In this section, we dive deeply into the procedure of TCD algorithm and optimize it dramatically by introducing an intuitive concept called \textit{tightest time interval} for temporal $k$-cores. In a nutshell, we directly prune subintervals without inducing their temporal $k$-cores if we can predict that the temporal $k$-cores are identical to other induced temporal $k$-cores, and tightest time interval is the key to fulfill prediction. In this way, the optimized TCD algorithm only performs TCD operations that are necessary for returning all distinct answers to a given TCQ instance. Conceptually, the new pruning operation of optimized algorithm eliminates the ``inter-core'' redundant computation, and the original TCD operation eliminates the ``intra-core'' redundant computation. Thus, the computational complexity of optimized algorithm no longer depends on the span of query time interval $[Ts,Te]$ like the baseline algorithm and the original TCD algorithm but only depends on the scale of final results.
	
Next, we introduce the concept and properties of tightest time interval in Section~\ref{sec:tti}, present three pruning rules based on tightest time interval for TCD algorithm in Section~\ref{sec:rules}, and briefly conclude and discuss the optimized TCD algorithm in Section~\ref{sec:otcd}.
	
\subsection{Tightest Time Interval (TTI)}\label{sec:tti}
	
We have such an observation, a temporal $k$-core of $[ts,te]$ may only contain edges with timestamps in a subinterval $[ts',te'] \subset [ts,te]$, since the edges in $[ts,ts')$ and $(te',te]$ have been removed by core decomposition. For example, consider a temporal $k$-core $\mathcal{T}_{[4,8]}^{2}$ illustrated in Figure~\ref{fig:tcdprocedure}. We can see that it does not contain edges with timestamps 4, 7 and 8. As a result, if we continue to induce $\mathcal{T}_{[4,7]}^{2}$ from $\mathcal{T}_{[4,8]}^{2}$ and to induce $\mathcal{T}_{[4,6]}^{2}$ from $\mathcal{T}_{[4,7]}^{2}$, the returned temporal $k$-cores remain unchanged. The sameness of temporal $k$-cores induced by different subintervals inspires us to further optimize TCD algorithm by pruning subintervals directly. As illustrated in Figure~\ref{fig:tcdprocedure}, the subintervals such as [4,7], [4,6], [5,8], [5,7] and [5,6] all induce the identical temporal $k$-cores to [4,8], so that they can be potentially pruned in advance.
	
For that, we propose the concept of Tightest Time Interval (TTI) for temporal $k$-cores. Given a temporal $k$-core of $[ts,te]$, its TTI refers to the minimal time interval $[ts',te']$ that can induce an identical temporal $k$-core to $\mathcal{T}^k_{[ts,te]}$, namely, there is no subinterval of $[ts',te']$ that can induce an identical temporal $k$-core to $\mathcal{T}^k_{[ts,te]}$. We formalize the definition of TTI as follows.
	
	\begin{definition}[Tightest Time Interval]\label{def:tti}
		Given a temporal $k$-core $\mathcal{T}_{[ts,te]}^{k}$, its tightest time interval $\mathcal{T}_{[ts,te]}^{k}$.TTI is $[ts',te']$, if and only if\\
		1) $\mathcal{T}_{[ts',te']}^{k}$ is an identical temporal $k$-core to $\mathcal{T}_{[ts,te]}^{k}$;\\
		2) there does not exist $[ts'',te''] \subset [ts',te']$, such that $\mathcal{T}_{[ts'',te'']}^{k}$ is an identical temporal $k$-core to $\mathcal{T}_{[ts,te]}^{k}$.
	\end{definition}
	
It is easy to prove the TTI of a temporal $k$-core of $[ts,te]$ is surely a subinterval of $[ts,te]$. To evaluate the TTI of a given $\mathcal{T}^k_{[ts,te]}$, we have the following theorem.
	
	\begin{theorem}\label{thm:eval}
		Given a temporal $k$-core $\mathcal{T}_{[ts,te]}^{k}$, $\mathcal{T}_{[ts,te]}^{k}$.TTI = $[t_{min}, t_{max}]$, where $t_{min}$ and $t_{max}$ are the minimum and maximum timestamps in $\mathcal{T}_{[ts,te]}^{k}$ respectively.
	\end{theorem}
	
	\begin{proof}
		On one hand, $\mathcal{T}^k_{[t_{min},t_{max}]}$ is identical to $\mathcal{T}_{[ts,te]}^{k}$. Because we can induce $\mathcal{T}^k_{[t_{min},t_{max}]}$ by TCD operation from $\mathcal{T}_{[ts,te]}^{k}$ due to $[t_{min},t_{max}] \subseteq [ts,te]$. Meanwhile, during the operation, none edge is actually removed since there is no edge with timestamp outsides $[t_{min},t_{max}]$ in $\mathcal{T}_{[ts,te]}^{k}$, and thus the temporal $k$-core $\mathcal{T}_{[ts,te]}^{k}$ will remain unchanged. On the other hand, any time interval $[ts',te'] \subset [t_{min},t_{max}]$ cannot induce a temporal $k$ core that is identical to $\mathcal{T}_{[ts,te]}^{k}$, since the edges with timestamp either $t_{min}$ or $t_{max}$ in $\mathcal{T}_{[ts,te]}^{k}$ are excluded at least.
	\end{proof}

With Theorem~\ref{thm:eval}, we can evaluate the TTI of a given temporal $k$-core instantly (by $O(1)$ time, see Section~\ref{sec:otcdimpl}), which guarantees the following optimization based on TTI will not incur extra overheads.
	
Moreover, there are the following important properties of TTI that support our pruning strategies.
	
	\begin{property}[Uniqueness]\label{theo:uniqueness}
		Given a temporal $k$-core $\mathcal{T}_{[ts,te]}^{k}$, there exists no other time interval than $\mathcal{T}_{[ts,te]}^{k}$.TTI evaluated by Theorem~\ref{thm:eval} that is also a TTI of $\mathcal{T}_{[ts,te]}^{k}$.
	\end{property}
	
	\begin{proof}
		Let $\mathcal{T}_{[ts,te]}^{k}$.TTI be $[ts',te']$, and $[ts'',te''] \neq [ts',te']$ be any other time interval. There are only two possibilities. Firstly, $[ts',te'] \not\subset [ts'',te'']$. However, the edges with timestamp $ts'$ and $te'$ are contained by $\mathcal{T}_{[ts,te]}^{k}$ according to Theorem~\ref{thm:eval}, and thereby $[ts'',te'']$ that does not cover $[ts',te']$ cannot induce $\mathcal{T}_{[ts,te]}^{k}$. Thus, the first possibility does not satisfy the first condition in Definition~\ref{def:tti}. Secondly, $[ts',te'] \subset [ts'',te'']$. However, since $[ts',te']$ can induce $\mathcal{T}_{[ts,te]}^{k}$, $[ts'',te'']$ is certainly not the tightest even if it can also induce $\mathcal{T}_{[ts,te]}^{k}$. Thus, the second possibility does not satisfy the second condition in Definition~\ref{def:tti}. Consequently, $[ts'',te''] \neq [ts',te']$ is not a TTI of $\mathcal{T}_{[ts,te]}^{k}$.
	\end{proof}
	
	\begin{property}[Equivalence]\label{thm:evaluivalence}
		Given two temporal $k$-cores $\mathcal{T}_{[ts,te]}^{k}$ and $\mathcal{T}_{[ts',te']}^{k}$, they are identical temporal graphs if and only if $\mathcal{T}_{[ts,te]}^{k}$.TTI = $\mathcal{T}_{[ts',te']}^{k}$.TTI.
	\end{property}
	
	\begin{proof}
		If $\mathcal{T}_{[ts,te]}^{k}$.TTI = $\mathcal{T}_{[ts',te']}^{k}$.TTI, $\mathcal{T}_{[ts,te]}^{k}$ and $\mathcal{T}_{[ts',te']}^{k}$ are both identical to the temporal $k$-core of the TTI according to Definition~\ref{def:tti}, and thus are identical to each other. Conversely, if $\mathcal{T}_{[ts,te]}^{k}$ and $\mathcal{T}_{[ts',te']}^{k}$ are identical, they must have a same unique TTI according to Theorem~\ref{thm:eval} and Property~\ref{theo:uniqueness}.
	\end{proof}
	
	\begin{property}[Inclusion]\label{theo:inclusion}
		Given two temporal $k$-cores $\mathcal{T}_{[ts,te]}^{k}$ and $\mathcal{T}_{[ts',te']}^{k}$, we have $\mathcal{T}_{[ts,te]}^{k}$.TTI $\subseteq$ $\mathcal{T}_{[ts',te']}^{k}$.TTI, if $[ts,te] \subseteq [ts',te']$.
	\end{property}
	
	\begin{proof}
		Since $[ts,te] \subseteq [ts',te']$, we have $\mathcal{T}_{[ts,te]}^{k}$ is a subgraph of $\mathcal{T}_{[ts',te']}^{k}$ according to Lemma~\ref{thm:subgraph}. Thus, the minimum timestamp in $\mathcal{T}_{[ts,te]}^{k}$ is certainly no earlier than the the minimum timestamp in $\mathcal{T}_{[ts',te']}^{k}$, and the maximum timestamp in $\mathcal{T}_{[ts,te]}^{k}$ is certainly no later than the the maximum timestamp in $\mathcal{T}_{[ts',te']}^{k}$. Then, according to Theorem~\ref{thm:eval}, we have $\mathcal{T}_{[ts,te]}^{k}$.TTI $\subseteq$ $\mathcal{T}_{[ts',te']}^{k}$.TTI.
	\end{proof}
	
	\begin{figure*}
		\centering
		\subfloat[Without pruning.]{\label{subfig:withoutpruning}
			\includegraphics[width=0.34\textwidth]{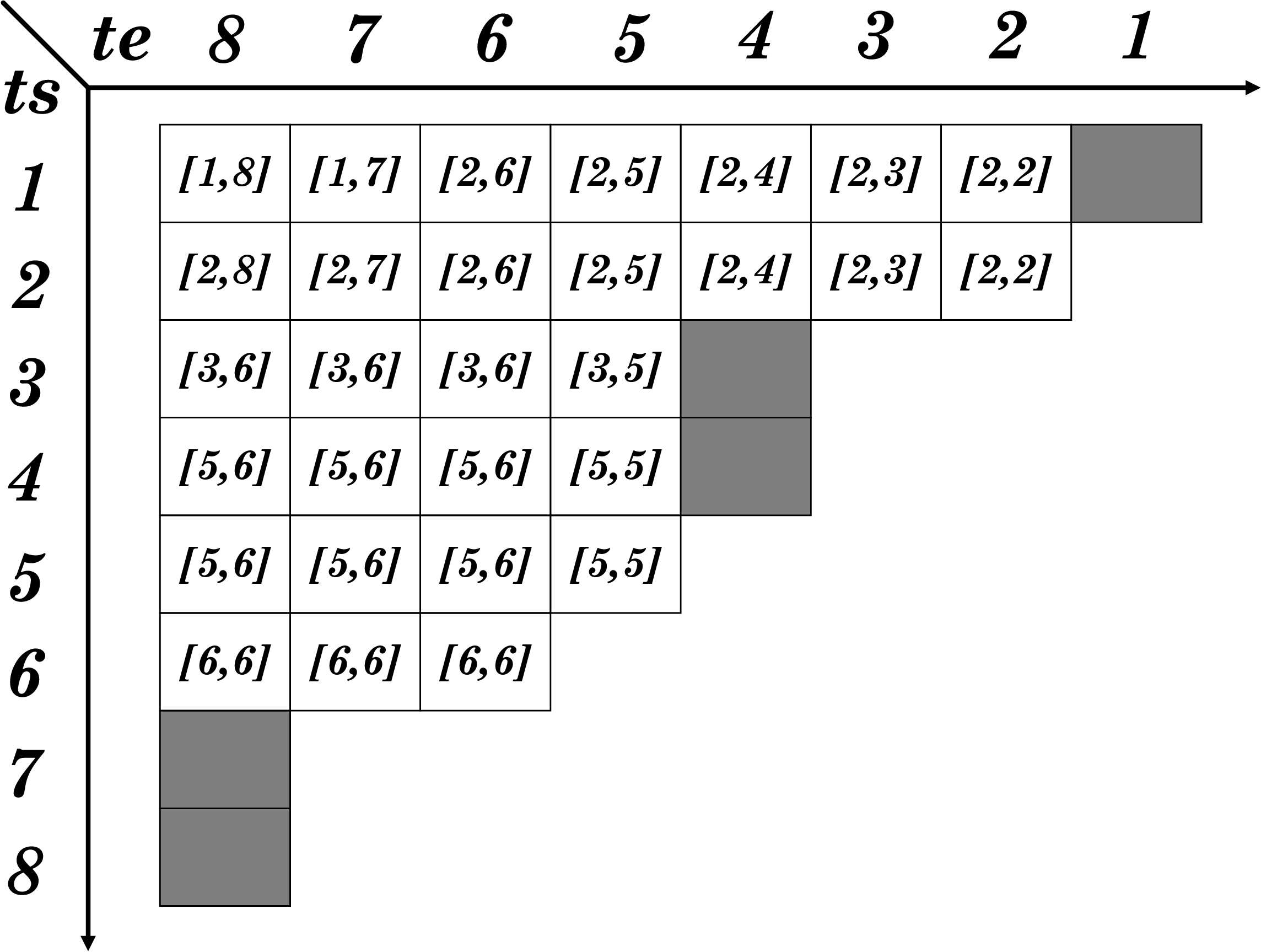}}
		\subfloat[With pruning.]{\label{subfig:withpruning}
			\includegraphics[width=0.66\textwidth]{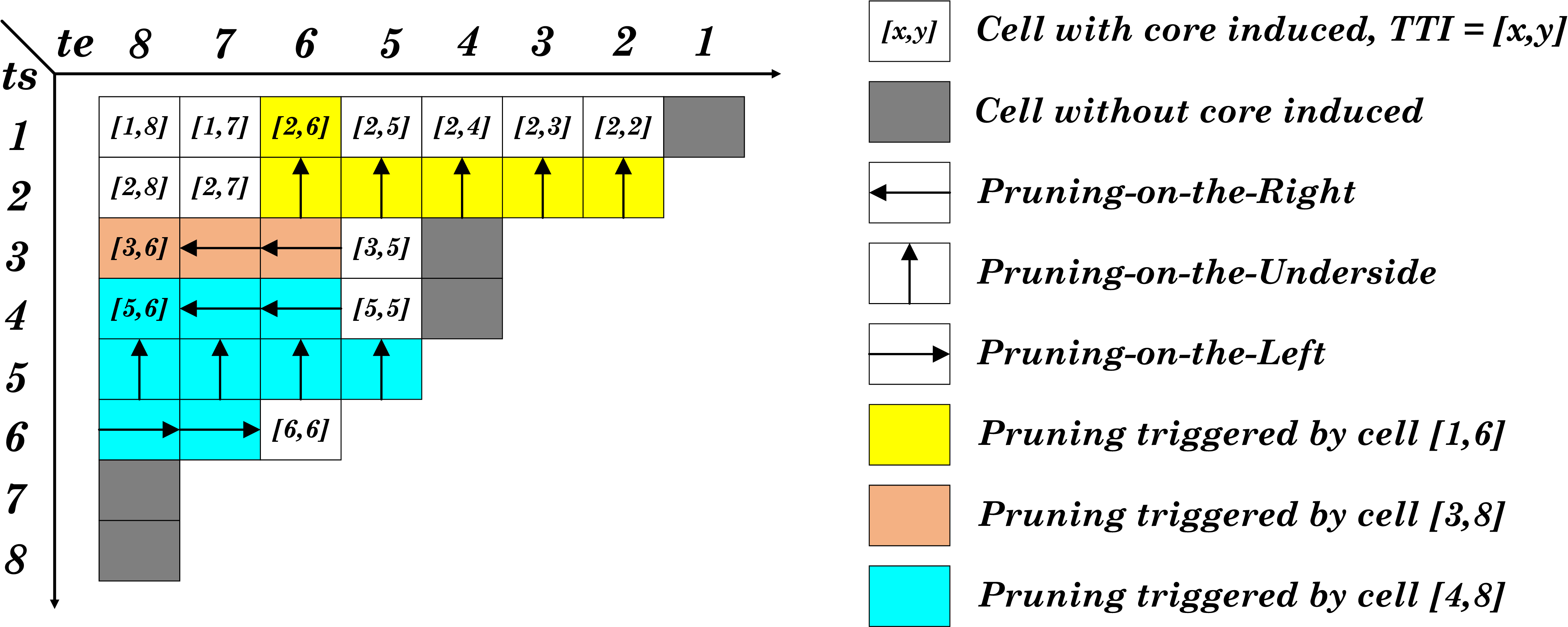}}
		\caption{Examples of subinterval pruning based on tightest time interval.}\label{fig:tcdproceduretil}
	\end{figure*}
	
Figure~\ref{subfig:withoutpruning} abstracts Figure~\ref{fig:tcdprocedure} as a schedule table of subinterval enumeration, and TCD algorithm will traverse the cells row by row and from left to right. For example, the cell in row 1 and column 6 represents a subinterval $[1,6]$, in which $[2,6]$ is the TTI of $\mathcal{T}^2_{[1,6]}$. In particular, the grey cells indicate that the temporal $k$-cores of the corresponding subintervals do not exist. Figure~\ref{subfig:withoutpruning} clearly reveals that TCD algorithm suffers from inducing a number of identical temporal $k$-cores (with the same TTIs). For example, the TTI $[5,6]$ repeats six times, which means six cells will induce identical temporal $k$-cores.

\subsection{Pruning Rules}\label{sec:rules}
	
The main idea of optimizing TCD algorithm is to predict the induction of identical temporal $k$-cores by leveraging TTI, thereby skipping the corresponding subintervals during the enumeration. Specifically, whenever a temporal $k$-core of $[ts,te]$ is induced, we evaluate its TTI $[ts',te']$. If $ts'>ts$ or/and $te'<te$, it is triggered that a number of subintervals on the schedule can be pruned in advance. According to different relations between $[ts,te]$ and $[ts',te']$, our pruning technique can be categorized into three rules which are not mutually exclusive. In other words, the three rules may be triggered at the same time, and prune different subintervals respectively. Next, we present these pruning rules in Section~\ref{sec:pr1}, Section~\ref{sec:pr2} and Section~\ref{sec:pr3}, respectively.
	
\subsubsection{Rule 1: Pruning-on-the-Right}\label{sec:pr1}
	
Consider the schedule illustrated in Figure~\ref{subfig:withoutpruning}. For each row, TCD algorithm traverses the cells (namely, subintervals) from left to right. If the TTI $[ts',te']$ in the current cell $[ts,te]$ meets such a condition, namely, $te'<te$, a pruning operation will be triggered, and the following cells in this row from $[ts,te-1]$ until $[ts,te']$ will be skipped because these subintervals will induce identical temporal $k$-cores to $\mathcal{T}^k_{[ts,te]}$. Since the pruned cells are on the right of trigger cell, we call this rule Pruning-On-the-Right (PoR). The pseudo code of PoR is given in lines 2-4 of Algorithm~\ref{alg:prune}. The correctness of PoR is guaranteed by the following lemma.
	
	\begin{lemma}\label{lem:piw}
		Given a temporal $k$-core $\mathcal{T}_{[ts,te]}^{k}$ whose TTI is $[ts',te']$, for any time interval $[ts,te'']$ with $te''\in [te',te]$, $\mathcal{T}_{[ts,te'']}^{k}$.TTI = $[ts',te']$. 
	\end{lemma}
	
	\begin{proof}
		On one hand, since $[ts,te''] \subseteq [ts,te]$, $\mathcal{T}_{[ts,te'']}^{k}$.TTI $\subseteq \mathcal{T}_{[ts,te]}^{k}$.TTI $= [ts',te']$ according to Inclusion (Property~\ref{theo:inclusion}). On the other hand, we can prove $[ts',te'] \subseteq \mathcal{T}_{[ts,te'']}^{k}$.TTI. If we induce $\mathcal{T}_{[ts',te']}^{k}$ from $\mathcal{T}_{[ts,te]}^{k}$ by TCD operation, it is easy to know $\mathcal{T}_{[ts,te]}^{k}$ will remain unchanged, because it only contains the edges with timestamps in $[ts',te']$ according to Theorem~\ref{thm:eval}. Thus, we have $\mathcal{T}_{[ts',te']}^{k}$.TTI = $[ts',te']$ according to Equivalence (Property~\ref{thm:evaluivalence}). Also, since $[ts',te'] \subseteq [ts,te'']$, $[ts',te']$ = $\mathcal{T}_{[ts',te']}^{k}$.TTI $\subseteq \mathcal{T}_{[ts,te'']}^{k}$.TTI according to Inclusion (Property~\ref{theo:inclusion}).
	\end{proof}
	
With Lemma~\ref{lem:piw}, we can predict that the TTIs in the cells $[ts,te-1]$, $\cdots$, $[ts,te']$ are the same as the trigger cell $[ts,te]$, when the PoR rule is satisfied. Thus, the temporal $k$-cores induced by these subintervals are all identical to the induced $\mathcal{T}^k_{[ts,te]}$ according to Equivalence (Property~\ref{thm:evaluivalence}).
	
For example, Figure~\ref{subfig:withpruning} illustrates two instances of PoR (the cells in orange and blue colors with left arrow). When $\mathcal{T}_{[3,8]}^{2}$ has been induced, we evaluate its TTI as $[3,6]$, and thus PoR is triggered. PoR immediately excludes the following two cells $[3,7]$ and $[3,6]$ from the schedule. As a proof, we can see the TTIs in these two cells are both $[3,6]$ in Figure~\ref{subfig:withoutpruning}.   
	
\subsubsection{Rule 2: Pruning-on-the-Underside}\label{sec:pr2}
	
We now consider $ts'>ts$, which causes pruning in the following rows but not the current row. So we call this rule Pruning-On-the-Underside (PoU). Specifically, if $ts'> ts$, for each row $r\in [ts+1,ts']$, the cells $[r,te]$, $[r,te-1]$, $\cdots$, $[r,r]$ will be skipped. The pseudo code of PoU is given in lines 5-8 of Algorithm~\ref{alg:prune}. The correctness of PoU is guaranteed by the following lemmas.
	
	\begin{lemma}\label{lem:pir2}
		Given a temporal $k$-core $\mathcal{T}_{[ts,te]}^{k}$ whose TTI is $[ts',te']$, for any time interval $[ts'',te]$ with $ts''\in [ts,ts']$, we have the TTI of $\mathcal{T}_{[ts'',te]}^{k}$ is $[ts',te']$. 
	\end{lemma}
	
	\begin{proof}
		The proof of this lemma is similar to Lemma~\ref{lem:piw} and thus is omitted.
	\end{proof}
	
	\begin{lemma}\label{lem:por1}
		Given a temporal $k$-core $\mathcal{T}_{[ts,te]}^{k}$ whose TTI is $[ts',te']$, for any time interval $[r,c]$ with $r\in[ts+1,ts']$ and $c\in [ts,te]$, we have $\mathcal{T}^k_{[r,c]}$ is identical to $\mathcal{T}^k_{[ts,c]}$.
	\end{lemma}
	
	\begin{proof}
		For $r\in[ts+1,ts']$, we have $\mathcal{T}_{[r,te]}^{k}$.TTI = $[ts',te']$ according to Lemma~\ref{lem:pir2}. Thus, $\mathcal{T}_{[r,te]}^{k}$ is identical to $\mathcal{T}_{[ts,te]}^{k}$ according to the Equivalence (Property~\ref{thm:evaluivalence}). Then, we have $\mathcal{T}_{[r,c]}^{k}$ is identical to $\mathcal{T}_{[ts,c]}^{k}$ when $c=te-1$ since them are induced by the same TCD operation from identical temporal graphs, and so on for the rest $[r,c]$ with the decrease of $c$ until $c = ts$.
	\end{proof}
	
Lemma~\ref{lem:por1} indicates that, PoU safely prunes some cells in the following rows, since these cells contain the same TTIs as their upper cells, which even have not been enumerated yet except the trigger cell. For example, Figure~\ref{subfig:withpruning} illustrates two PoU instances (the cells in yellow and blue colors with up arrow). On enumerating the cell $[1,6]$, since the contained TTI is $[2,6]$, the cells $[2,6]$, $\cdots$, $[2,2]$ are pruned by PoU, because the TTIs in these cells are the same as the cells $[1,6]$, $\cdots$, $[1,2]$ respectively, though the TTIs of cells $[1,5]$, $\cdots$, $[1,2]$ have not been evaluated.
	
\subsubsection{Rule 3: Pruning-on-the-Left}\label{sec:pr3}
	
Lastly, if both $ts'> ts$ and $te' < te$, for each row $r\in [ts'+1,te']$, the cells $[r,te]$, $[r,te-1]$, $\cdots$, $[r,te'+1]$ will also be skipped, besides the cells pruned by PoR and PoU. Although these cells are in the rows under the current row $ts$, the temporal $k$-core of each of them is identical to the temporal $k$-core of a cell (namely, $[r,te']$) on the right in the same row but not its upper cell like PoU. So we call this rule Pruning-On-the-Left (PoL). The pseudo code of PoL is given in lines 9-12 of Algorithm~\ref{alg:prune}. The correctness of PoL is guaranteed by the following lemma.
	
	\begin{lemma}\label{lem:por2}
		Given a temporal $k$-core $\mathcal{T}_{[ts,te]}^{k}$ whose TTI is $[ts',te']$, for any time interval $[r,c]$ with $r\in[ts'+1,te']$ and $c\in [te'+1,te]$, we have $\mathcal{T}^k_{[r,c]}$ is identical to $\mathcal{T}^k_{[r,te']}$.
	\end{lemma}
	
	\begin{proof}
		Assume $\mathcal{T}^k_{[r,c]}$.TTI = $[r',c']$. According to Inclusion (Property~\ref{theo:inclusion}), we have $[r',c'] \subseteq [ts',te']$ since $[r,c] \subseteq [ts,te]$. Thus, $c' \leqslant te'$. Then, according to Lemma~\ref{lem:piw}, we have $\mathcal{T}^k_{[r,te']}$.TTI = $[r',c']$ since $te' \in [c',c]$. Lastly, according to Equivalence (Property~\ref{thm:evaluivalence}), we have $\mathcal{T}^k_{[r,c]}$ is identical to $\mathcal{T}^k_{[r,te']}$.
	\end{proof}
	
For example, Figure~\ref{subfig:withpruning} illustrates a PoL instance (the cells in blue color with right arrow). On enumerating the cell $[4,8]$, PoL is triggered since the contained TTI is $[5,6]$. Then, the cells $[6,8]$ and $[6,7]$ are pruned by PoL because the TTIs contained in them are the same as the cell $[6,6]$ on the right of them. PoL is more tricky than PoU because the cells are pruned for containing the same TTIs as other cells that are scheduled to traverse after them by TCD algorithm. Note that, the cell $[4,8]$ triggers all three kinds of pruning. In fact, a cell may trigger PoL only, PoU only, or all three rules.
	
\subsection{Optimized TCD Algorithm}\label{sec:otcd}
	
Compared with TCD algorithm, the improvement of Optimized TCD (OTCD) algorithm is simply to conduct a pruning operation whenever a temporal $k$-core has been induced. Specifically, we evaluate the TTI of this temporal $k$-core, check each pruning rule to determine if it is triggered, and prune the specific subintervals on the schedule in advance. The pseudo code of pruning operation is given in Algorithm~\ref{alg:prune}. Note that, the ``prune'' in Algorithm~\ref{alg:prune} is a logical concept, and can have different physical implementations.
	
	\begin{algorithm}[t!]
		\DontPrintSemicolon
		\KwIn{$[ts,te]$ and $\mathcal{T}^k_{[ts,te]}$}
		$[ts',te'] \leftarrow \mathcal{T}^k_{[ts,te]}$.TTI \tcp*[h]{Theorem~\ref{thm:eval}}\;
		\If(\tcp*[f]{Rule 1: PoR}){$te' < te$}{
		    \For{$c$ $\leftarrow$ $te$ - 1 \textbf{to} $te'$}{
		        prune the subinterval $[ts,c]$\;
		    }
		}
		\If(\tcp*[f]{Rule 2: PoU}){$ts' > ts$ }{
		    \For{$r$ $\leftarrow$ $ts + 1$ \textbf{to} $ts'$}{
		        \For{$c$ $\leftarrow$ te \textbf{to} r}{
		            prune the subinterval $[r,c]$\;
		        }
		    }
		}
		\If(\tcp*[f]{Rule 3: PoL}){$ts' > ts$ and $te' < te$}{
		    \For{r $\leftarrow$ ts'+1 \textbf{to} te'}{
		        \For{c $\leftarrow$ te \textbf{to} te'+1}{
		            prune the subinterval $[r,c]$\;
		        }
		    }
		}
		\caption{Pruning operation.}\label{alg:prune}
	\end{algorithm}
	
As illustrated in Figure~\ref{subfig:withpruning}, OTCD algorithm completely eliminates repeated inducing of identical temporal $k$-cores, namely, each distinct temporal $k$-core is induced exactly once during the whole procedure. It means, the real computational complexity of OTCD algorithm is the summation of complexity for inducing each distinct temporal $k$-core but not the temporal $k$-core of each subinterval of $[Ts, Te]$. Therefore, we say OTCD algorithm is scalable with respect to the query time interval $[Ts, Te]$. For many real-world datasets, the span of $[Ts, Te]$ could be very large, while there exist only a limited number of distinct temporal $k$-cores over this period, so that OTCD algorithm can still process the query efficiently.
	
\section{Implementation}\label{sec:otcdimpl}
	
In this section, we address the physical implementation of proposed algorithm. We first introduce a data structure for temporal graph representation in Section~\ref{sec:tel}, based on which we explain the details of TCD Operation implementation in Section~\ref{sec:tcdimpl}.
	
\subsection{Temporal Edge List (TEL)}\label{sec:tel}
	
\begin{figure*}
		\centering
		\includegraphics[width=0.9\textwidth]{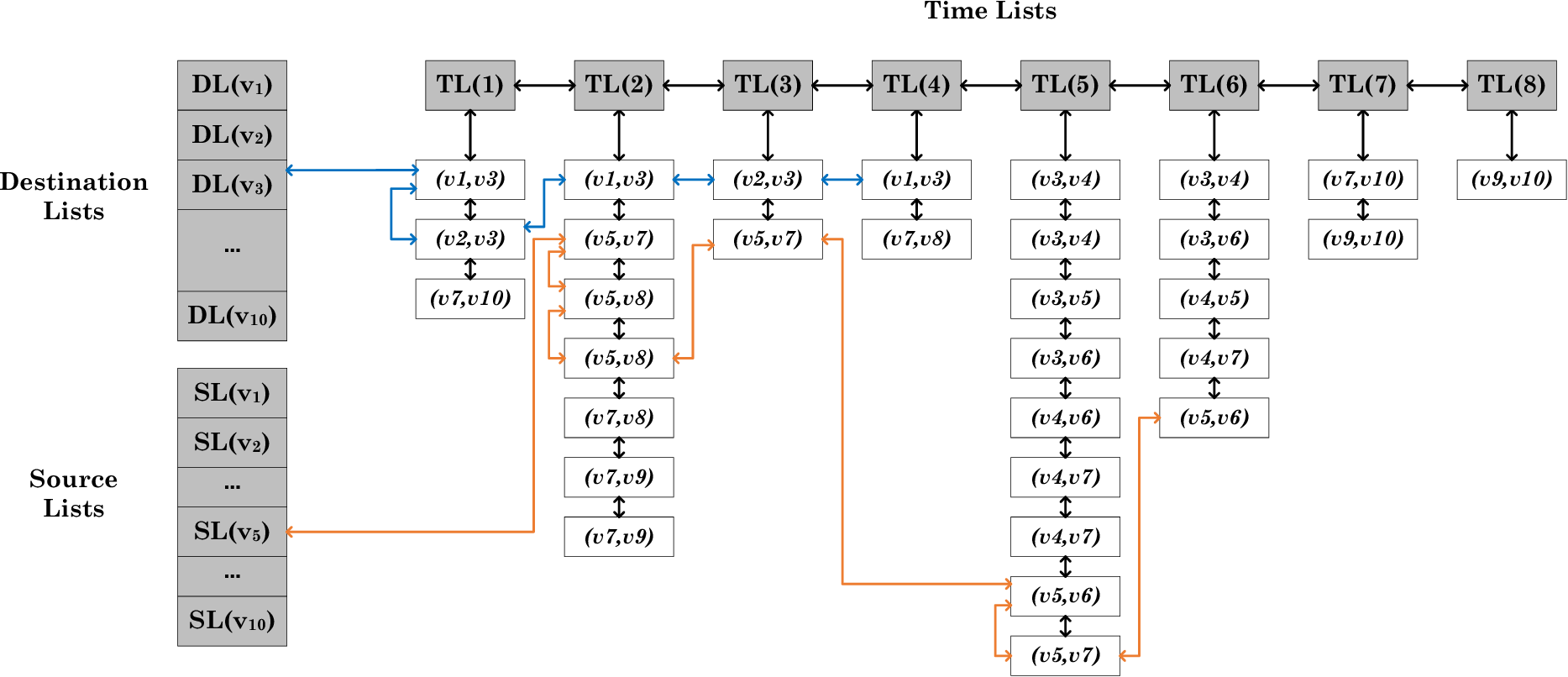}
		\caption{The conceptual illustration of a partial TEL of our running example graph.}\label{fig:telultra}
	\end{figure*}

We propose a novel data structure called Temporal Edge List (TEL) for representing an arbitrary temporal graph (including temporal $k$-cores that are also temporal graphs), which is both the input and output of TCD operation. Conceptually, TEL($\mathcal{G}$) preserves a temporal graph $\mathcal{G}=(\mathcal{V},\mathcal{E})$ by organizing its edges in a 3-dimension space, each dimension of which is a set of bidirectional linked lists, as illustrated in Figure~\ref{fig:telultra}. The first dimension is time, namely, all edges in $\mathcal{E}$ are grouped by their timestamps. Each group is stored as a bidirectional linked list called Time List (TL), and TL($t$) denotes the list of edges with a timestamp $t$. Then, TEL($\mathcal{G}$) uses a bidirectional linked list, in which each node represents a timestamp in $\mathcal{G}$, as a timeline in ascending order to link all TLs, so that some temporal operations can be facilitated. Moreover, the other two dimensions are source vertex and destination vertex respectively. We use a container to store the Source Lists (SL) or Destination Lists (DL) for each vertex $v \in \mathcal{V}$, where SL($v$) or DL($v$) is a bidirectional linked list that links all edges whose source or destination vertex is $v$. Actually, an SL or DL is an adjacency list of the graph, by which we can retrieve the neighbor vertices and edges of a given vertex efficiently. Given a temporal graph $\mathcal{G}$, TEL($\mathcal{G}$) is built in memory by adding its edges iteratively. For each edge $(u, v, t) \in \mathcal{E}$, it is only stored once, and TL($t$), SL($u$) and DL($v$) will append its pointer at the tail respectively.
	
Figure~\ref{fig:telultra} illustrates a partial TEL of our example graph. The SLs and DLs other than SL($v_5$) and DL($v_3$) are omitted for conciseness. Basically, TL, SL and DL offer the functionality of retrieving edges by timestamp and linked vertex respectively. For example, for removing all neighbor edges of a vertex $v$ with degree less than $k$ in TCD operation, we can locate SL($v$) and DL($v$) to retrieve these edges. Moreover, the linked list of TL can offer efficient temporal operations. For example, for truncating $\mathcal{G}$ to $\mathcal{G}_{[ts,te]}$ in TCD operation, we can remove TL($t$) with $t<ts$ or $t>te$ from the linked list of TL conveniently. To get the TTI of a temporal $k$-core, we only need to check the head and tail nodes of the linked list of TL in its TEL to get the minimum and maximum timestamps respectively. The superiority of TEL is summarized as follows.
	\begin{itemize}
		\item By TCD operation, a TEL will be trimmed to a smaller TEL, and there is none intermediate TEL produced. Thus, the memory requirement of (O)TCD algorithm only depends on the size of initial TEL($\mathcal{G}_{[Ts,Te]}$).
		\item TEL consumes $O(|\mathcal{E}|)$ space for storing a temporal graph, which is optimal because at least $O(|\mathcal{E}|)$ space is required for storing a graph (e.g., adjacency lists). Although there are $6|\mathcal{E}| + 2|\mathcal{V}| + 3n$ pointers of TLs, SLs and DLs stored additionally, TEL is still compact compared with PHC-Index, where $n$ is the number of timestamps in the graph.
		\item TEL supports the basic manipulations listed in Table~\ref{table:telop} in constant time, which are cornerstones of implementing our algorithms and optimization techniques.
		\item For dynamic graphs, when a new edge coming, TEL simply appends a new node representing the current time at the end of linked list of TL, and then adds this edge as normal. Thus, TEL can also deal with dynamic graphs.
	\end{itemize}

\begin{table*}[t]
    %\begin{threeparttable}[t]
	\centering
	\caption{The basic manipulations of TEL.}\label{table:telop}
	\small
	\begin{tabular}{clc}
		\hline
		Name & Functionality & Complexity\\\hline
		next\_TL($TL$) / prev\_TL($TL$) & get the next or previous TL in the linked list of TL & $O(1)$ \\
		get\_SL($v$) / get\_DL($v$) & get the SL or DL of a given vertex $v$ from a hash map & $O(1)$ \\
		del\_TL($TL$) & remove the given TL node from the linked list of TL & $O(1)$ \\
		del\_edge($e$) & delete a given edge $e = (u,v,t)$ and update TL($t$), SL($u$) and DL($v$) respectively & $O(1)$\\ 
		get\_TTI() & return the timestamps of head and tail nodes of linked list of TL & $O(1)$\\
		\hline
	\end{tabular}
	%begin{tablenotes}
	    %\item[1] $l$ is the offset of locating $t$ in the linked list of TL. For each fixed $ts$, the total cost of get\_TL() is at most $Te-ts$ in our algorithm.
	%\end{tablenotes}
	%\end{threeparttable}
\end{table*}

\subsection{Implement TCD Operation on TEL}\label{sec:tcdimpl}

Given a TCQ instance, our algorithm starts to work on a copy of TEL($\mathcal{G}_{[Ts,Te]}$) in memory, which is obtained by truncating TEL($\mathcal{G}$). Then, our algorithm only needs to maintain an instance of TEL($\mathcal{T}^k_{[ts,Te]}$) and another instance of TEL($\mathcal{T}^k_{[ts,te]}$) with $[ts,te]$ $\subseteq$ $[Ts,Te]$ in memory. The first instance is used to induce the first temporal $k$-core $\mathcal{T}^k_{[ts+1,Te]}$ by TCD for each row in Figure~\ref{fig:tcdprocedure}. The second instance is used to induce the following temporal $k$-cores $\mathcal{T}^k_{[ts,te-1]}$ by TCD in each row. Each TCD operation is decomposed to a series of TEL manipulations, and trims the input TEL without producing any intermediate data.

\begin{comment}
To assist the implementation of TCD operation, the following global data structures are used by our algorithm. 1) Vertex degree map $\mathbb{M}_{v}$ preserves the number of neighbor for each vertex in the maintained TEL, and provides the ability of immediately obtaining the degree of a given vertex. 2) Edge count map $\mathbb{M}_{e}$  preserves the number of parallel edges between each pair of linked vertices in the maintained TEL, and allows for maintaining the connectivity between vertices efficiently. 3) Vertex degree heap $\mathbb{H}_{v}$ organizes all vertices in the maintained TEL into a minimum heap ordered by their degree, so that the vertices with less than $k$ neighbors can be retrieved directly.    
\end{comment}	

To assist the implementation of TCD operation, our algorithm uses a global data structure $\mathbb{H}_{v}$ that organizes all vertices in the maintained TEL into a minimum heap ordered by their degree, so that the vertices with less than $k$ neighbors can be retrieved directly. Note that, whenever an edge is deleted from the maintained TEL, $\mathbb{H}_{v}$ will also be updated due to the possible decrease of vertex degrees. The trivial details of updating $\mathbb{H}_{v}$ is omitted.

\begin{comment}
\begin{itemize}
	\item Vertex Degree Map $\mathbb{M}_{v}$: $v\in \mathcal{V}_{[Ts,Te]} \mapsto |\{u|u\in \mathcal{V}_T, (v,u,t)\in \mathcal{E}_T\}| \geqslant 1$, where $\mathcal{V}_{T}$ is the set of vertices and $\mathcal{E}_T$ is the set of edges in the maintained TEL respectively.
	\item Edge Count Map $\mathbb{M}_{e}$: $(v\in \mathcal{V}_{T}, u\in \mathcal{V}_{T}) \mapsto |\{(v,u,t) | (v,u,t)\in \mathcal{E}_T\}|$.
	\item Vertex Minimum Heap $\mathbb{H}_{v}$ ordered by the degree of vertex.
\end{itemize}    
\end{comment}

	\begin{algorithm}[t]
		\DontPrintSemicolon
		\KwIn{TEL($\mathcal{G}$), $[ts,te]$, $k$}
		\KwOut{TEL($\mathcal{T}_{[ts,te]}^{k}$)}
		$TL$ $\leftarrow$ the head of linked list of TL in TEL($\mathcal{G}$)\;
		\While{$TL$.timestamp $\neq$ $ts$}{
			\For{edge $e$ in $TL$}{
				del\_edge($e$)\;
				udpate $\mathbb{H}_{v}$\;
				}
			del\_TL($TL$)\;
			$TL$ $\leftarrow$ next\_TL($TL$)\;
			%TEL($\mathcal{G}$).delete\_node(node.next\_node)\;
		}
		$TL$ $\leftarrow$ the tail of linked list of TL in TEL($\mathcal{G}$)\;
		\While{$TL$.timestamp $\neq$ $te$}{
			\For{edge $e$ in $TL$}{
				del\_edge($e$)\;
				udpate $\mathbb{H}_{v}$\;
			}
			del\_TL($TL$)\;
			$TL$ $\leftarrow$ prev\_TL($TL$)\;
			%TEL($\mathcal{G}$).delete\_node(node.next\_node)\;
		}
		\While{$\mathbb{H}_v$ is not empty and $\mathbb{H}_v$.peek $<$ $k$}{
			vertex $v$ $\leftarrow$ $\mathbb{H}_{v}$.pop()\;
			\For{edge $e$ in SL($v$)}{
				del\_edge($e$)\;
				del\_TL(TL($e$.timestamp)) if the TL is empty\;
				update $\mathbb{H}_{v}$\;
			}
			\For{edge $e$ in DL($v$)}{
				del\_edge($e$)\;
				del\_TL(TL($e$.timestamp)) if the TL is empty\;
				update $\mathbb{H}_{v}$\;
			}
		}
		\caption{TCD operation in Algorithm~\ref{alg:tcd}}\label{alg:tcdop}
	\end{algorithm}
	
 Algortithm~\ref{alg:tcdop} gives the implementation of TCD operation on TEL. The algorithm takes as input the TEL of a given graph $\mathcal{G}$, along with the parameters $k$, $ts$ and $te$ specifying the target temporal $k$-core $\mathcal{T}_{[ts,te]}^{k}$. In truncation phase, TEL($\mathcal{G}$) is projected to TEL($\mathcal{G}_{[ts,te]}$) (lines 1-14). Specifically, the linked list of TL is traversed from the head and tail bidirectionally until meeting $ts$ and $te$ respectively. For each node representing the timestamp $t$ traversed, the edges in TL($t$) are removed from TEL, and $\mathbb{H}_{v}$ is updated for each edge removed. In decomposition phase, TEL($\mathcal{G}_{[ts,te]}$) is further transformed to TEL($\mathcal{T}_{[ts,te]}^{k}$) (lines 15-24). Specifically, the algorithm pops the vertex with the least neighbors from $\mathbb{H}_{v}$ iteratively until the remaining vertices all have at least $k$ neighbors or the heap is empty. For each popped vertex $v$, it removes the linked edges of $v$ preserved in SL($v$) and DL($v$) from TEL respectively and updates $\mathbb{H}_{v}$ accordingly. In particular, a TL will be removed from the linked list of TL after the last edge in it has been removed (lines 19 and 23).

\begin{comment}
	\begin{algorithm}
		\DontPrintSemicolon
		\KwIn{TEL($\mathcal{G}$), $k$, $[ts,te]$}
		\KwOut{TEL($\mathcal{T}_{[ts,te]}^{k}(\mathcal{G})$)}
		truncation($ts$, $te$)\;
		decomposition($k$)\;
		\SetKwProg{Fn}{Procedure}{}{}
		\Fn{truncation($ts$,$te$)}{
			\For{$e=(u,v,t)$ preserved in $\{TL(t)|t<ts||t>te\}$}{
				remove $e$ from TEL($\mathcal{G}$)\;
				decrease Mc[$(u,v)$] by 1\;
			}
		}
		\Fn{decomposition($k$)}{
			\While{minumum vertex degree in Mv is less than $k$}{
				pop $(deg(v), v)$ from Hv\;
				remove $v$ from Mv\;
				\For{$e=(u,v,t)$ preserved in $TS(u), TD(v)$}{
					remove $e$ from TEL($\mathcal{G}_{[ts,te]}$)\;
					remove $(u,v)$ from Mc\;
				} 
			}
		}
		\caption{TCD.}\label{alg:tcdop}
	\end{algorithm}
\end{comment}	
	
\begin{figure*}
		\centering
		\includegraphics[width=\textwidth]{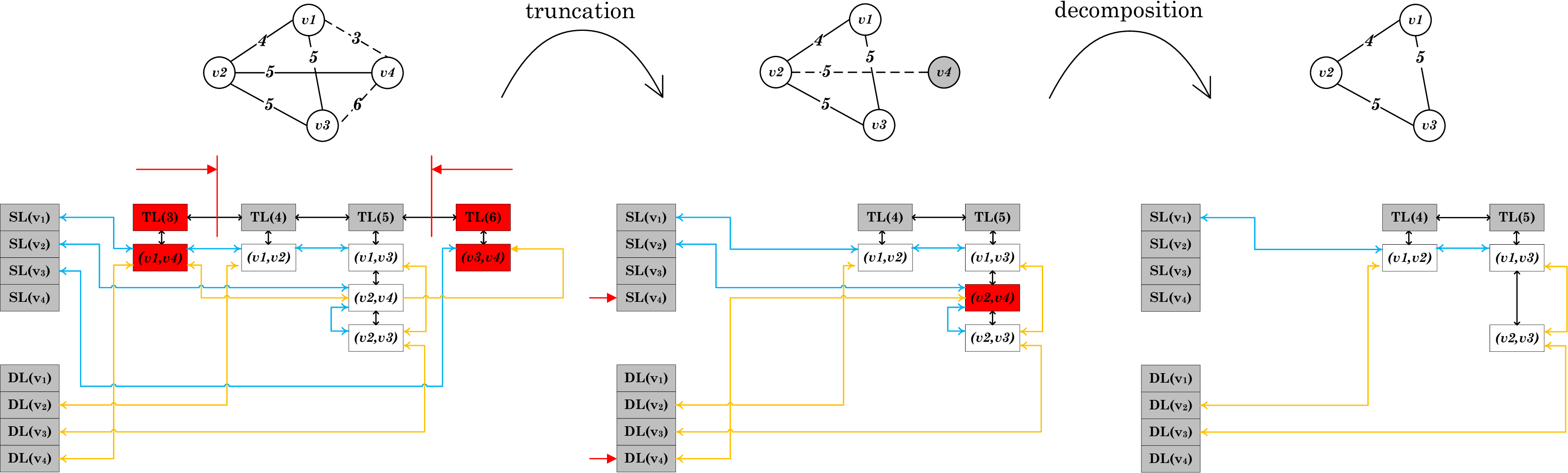}
		\caption{An example of TCD operation on TEL.}\label{fig:tcdop}
	\end{figure*}

To clarify the procedure of Algorithm~\ref{alg:tcdop}, Figure~\ref{fig:tcdop} illustrates an example of inducing $\mathcal{T}_{[4,5]}^{2}$ from $\mathcal{T}_{[3,6]}^{2}$. The edges are going to be deleted are marked in red color. We can see that, the procedure is actually a stream of edge deletion, while TEL maintains the entries to retrieve the remaining edges.

\subsection{Complexity}
\noindent
\textbf{TCD.} Theoretically, the complexity of TCD algorithm is bounded by $\sum_{t=Ts}^{Te}\{(|\mathcal{V}_{[t,Te]}|+|\mathcal{E}_{[t,Te]}|)\log|\mathcal{V}_{[t,Te]}|+m|\mathcal{E}_{[t,Te]}|\}$, where $m$ is a small constant. For each anchored $t$, TCD algorithm gradually peels $\mathcal{T}_{[t,Te]}^{k}$ like an onion by TCD operation until it contains none temporal $k$-core. In the process, there are at most $|\mathcal{E}_{[t,Te]}|$ edges deleted, and deleting each edge takes 
a small constant time $O(m)$ for TEL updating and at most $O(\log|\mathcal{V}_{[t,Te]}|)$ time for $\mathbb{H}_{v}$ maintenance. Similarly, there are at most $|\mathcal{V}_{[t,Te]}|$ vertices deleted, and deleting each vertex takes $O(\log|\mathcal{V}_{[t,Te]}|)$ time for $\mathbb{H}_{v}$ maintenance. Therefore, The total time overhead is the sum of edge and vertex deleting costs.

Note that, the complexity of TCD algorithm can also be represented by $O((Te-Ts)^2B)$ according to Algorithm~\ref{alg:tcd}, where $B$ is the average time overhead of TCD operation. However, $B$ cannot be estimated precisely, since each TCD operation may delete zero to $|\mathcal{E}_{[t,Te]}|$ edges. Therefore, we bound the complexity by the maximum deleting cost according to Algorithm~\ref{alg:tcdop}, which is more reasonable.

\noindent
\textbf{OTCD.} The complexity of OTCD algorithm is simply bounded by $\sum(|V^*|+|E^*|)\log|V^*| + m|E^*|$, where $V^*$ and $E^*$ refer to the sets of vertices and edges that have to be deleted for inducing the result temporal $k$-cores respectively. Due to the pruning rules, there are much less temporal $k$-cores induced by OTCD algorithm. Thus, $|V^*|$ and $|E^*|$ are orders of magnitude less than the total number of vertices and edges deleted in TCD algorithm, most of which are actually used for inducing identical temporal $k$-cores, though they cannot be really estimated.
	
\section{Extension}
\label{sec:ext}

To demonstrate the wide applicability of our approach in practice, we present several typical scenarios that extends the data model or query model of TCQ, and sketch how to address them based on our data structure and algorithm in this section. 

\subsection{Data Model Extension}\label{sec:extdata}

\textbf{Dynamic Graph.}
Since most real-world graphs are evolving over time, it is significant to fulfill TCQ on dynamic graphs. Benefitted from its design in ``timeline'' style, our data structure TEL can deal with new edges naturally in memory through two new manipulations add\_TL($t$) and add\_edge($u$, $v$, $t$). When a new edge $(u,v,t)$ arrived, we firstly create an empty TL($t$), and append it at the end of the linked list of TL since $t$ is obviously greater than the existing timestamps. Then, we create a new edge node for $(u,v,t)$ and append it to TL($t$), SL($u$) and DL($v$) respectively. Both manipulations are finished in constant time. The maintenance of a dynamic TEL is actually consistent with the construction of a static TEL. Therefore, our (O)TCD algorithm can run on the dynamic TEL anytime.

In contrast, updating PHC-Index is a non-trivial process. Although there are previous work~\cite{li2014dynamic, sar2016incremental} on coreness updating for dynamic graphs, the update is only valid for the whole life time of graph. While, for an arbitrary start time, it is uncertain whether the coreness of a vertex will be changed by a new edge.

\subsection{Query Model Extension}\label{sec:extquery}

The existing graph mining tasks regarding $k$-core introduce various constraints. For temporal graphs, we only focus on the temporal constraints. In the followings, we present two of them that can be integrated into TCQ model and also be addressed by our algorithm directly, which demonstrate the generality of our model and algorithm.

\textbf{Link Strength Constraint.}
In the context of temporal graph, link strength usually refers to the number of parallel edges between a pair of linked vertices. Obviously, the minimum link strength in a temporal $k$-core represents some important properties like validity, since noise interaction may appear over time and a pair of vertices with low link strength may only have occasional interaction during the time interval. Actually, the previous work~\cite{wu2015core} has studied this problem without the time interval constraint. Therefore, it is reasonable to extend TCQ to retrieve $k$-cores with a lower bound of link strength during a given time interval. It can be achieved by trivially modifying the TCD Operation. Specifically, the modified TCD Operation will remove the edges between two vertices once the number of parallel edges between them is decreased to be lower than the given lower bound of link strength, while the original TCD operation will do this when the number becomes zero. Thus, the modification brings almost none extra time and space consumption.

\textbf{Time Span Constraint.}
In many cases, we prefer to retrieve temporal $k$-cores with a short time span (between their earliest and latest timestamps), which is similar to the previous work on density-bursting subgraphs~\cite{chu2019online}. Because such a kind of short-term cohesive subgraphs tend to represent the occurrence of some special events. TCQ can be conveniently extended for resolving the problem by specifying a constraint of time span. Since the time span of a temporal $k$-core is preserved in its TEL, which is actually the length of its TTI, we can abandon the temporal $k$-cores returned by TCD operation that cannot satisfy the time span constraint on the fly. It brings almost no extra time and space consumption. Moreover, we can also extend TCQ to find the temporal $k$-core with the shortest or top-$n$ shortest time span.

\section{Experiment}
In this section, we conduct experiments to verify both efficiency and effectiveness of the proposed algorithm on a Windows machine with Intel Core i7 2.20GHz CPU and 64GB RAM. The algorithms are implemented through C++ Standard Template Library. Our source codes are shared on GitHub\footnotemark.

\footnotetext{https://github.com/ThomasYang-algo/Temporal-k-Core-Query-Project}
	
\subsection{Dataset}

We choose seven temporal graphs with different sizes and domains for our experiments. The first three graphs are obtained from KONECT Project~\cite{kunegis2013konect}, and the other four graphs are obtained from the SNAP~\cite{snapnets}. The basic statistics of these graphs are given in Table~\ref{table:dataset}. All timestamps are unified to integers in seconds.
	
	\begin{table}[t]
		\centering
		\caption{Datasets.}\label{table:dataset}
		\begin{tabular}{lccc}
			\hline
			Name & |$\mathcal{V}$| & |$\mathcal{E}$| & Span(days)\\
			\hline
			Youtube & 3.2M & 9.4M & 226\\
			DBLP & 1.8M & 29.5M & 17532\\
			Flickr & 2.3M & 33M & 198\\
			CollegeMsg & 1.8K & 20K & 193\\
			email-Eu-core-temporal & 0.9K & 332K & 803\\
			sx-mathoverflow & 24.8K & 506K & 2350\\
			sx-stackoverflow & 2.6M & 63.5M & 2774\\
			\hline
		\end{tabular}
	\end{table}

\subsection{Efficiency}
	\begin{table}
		\centering
		\caption{Selected temporal $k$-core queries.}\label{table:query}
		\begin{tabular}{lllllc}
			\hline
			id & $\mathcal{G}$ & $ts$ (sec) & $te$ (sec) & $k$ & result \#\\
			\hline
			1 & CollegeMsg & 554400 & 565200 & 2 & 61\\
			2 & CollegeMsg & 558000 & 568800 & 2 & 21\\
			3 & CollegeMsg & 561600 & 572400 & 2 & 27\\
			4 & CollegeMsg & 565200 & 576000 & 2 & 26\\
			5 & CollegeMsg & 568800 & 579600 & 2 & 10\\
			6 & email-Eu-core-temporal & 36000 & 46800 & 3 & 2\\
			7 & email-Eu-core-temporal & 39600 & 50400 & 3 & 3\\
			8 & email-Eu-core-temporal & 284400 & 295200 & 3 & 7\\
			9 & email-Eu-core-temporal & 288000 & 298800 & 3 & 25\\
			10 & email-Eu-core-temporal & 291600 & 302400 & 3 & 16\\
			11 & sx-mathoverflow & 864000 & 867600 & 2 & 8\\
			12 & sx-mathoverflow & 1116000 & 1119600 & 2 & 4\\
			13 & sx-mathoverflow & 1389600 & 1393200 & 2 & 5\\
			14 & sx-mathoverflow & 1483200 & 1486300 & 2 & 2\\
			15 & sx-mathoverflow & 1738800 & 1742400 & 2 & 8\\
			16 & sx-stackoverflow & 378000 & 381600 & 2 & 6\\
			17 & sx-stackoverflow & 417600 & 421200 & 2 & 37\\
			18 & sx-stackoverflow & 421200 & 424800 & 2 & 5\\
			19 & sx-stackoverflow & 424800 & 428400 & 2 & 5\\
			20 & sx-stackoverflow & 486000 & 489600 & 2 & 10\\
			\hline
		\end{tabular}
	\end{table}
	
		\begin{table}
		\centering
		\caption{Effect of pruning rules.}\label{table:prune}
		\begin{tabular}{c|ccc|ccc|c}
			\hline 
            \multirow{2}{*}{id} & 
            \multicolumn{3}{|c|}{Triggered Times} &
            \multicolumn{4}{|c}{Pruned Cell Percentage (\%)} \\
            \cline{2-8}
            &  PoR & PoU & PoL & PoR & PoU & PoL & Total\\
			\hline
			1 & 54 & 72 & 2 & 0.02 & 72 & 23.6 & 95.62\\
			6 & 2 & 4 & 1 & 0.01 & 51.8 & 32.1 & 83.91\\
			11 & 8 & 10 & 1 & 0.04 & 57.1 & 24.5 & 81.64\\
            16 & 5 & 9 & 1 & 0.04 & 56.9 & 33.5 & 90.44\\
			\hline
		\end{tabular}
	\end{table}

To evaluate the efficiency of our algorithm, we firstly manually select twenty temporal $k$-core queries from tested random queries with a time span (namely, $Te-Ts$) of 1-3 days, which have been verified to be valid, namely, there is at least one temporal $k$-core returned for each query. The setting of time span is moderate, otherwise other algorithms than OTCD can hardly stop successfully. Table~\ref{table:query} gives the details of query parameters, so that other researchers can reverify our experimental results or compare with our approach with the same queries.
	
\begin{figure}
		\centering
		\subfloat[CollegeMsg]{\label{subfig:collegemsg-col}
			\includegraphics[width=0.5\linewidth]{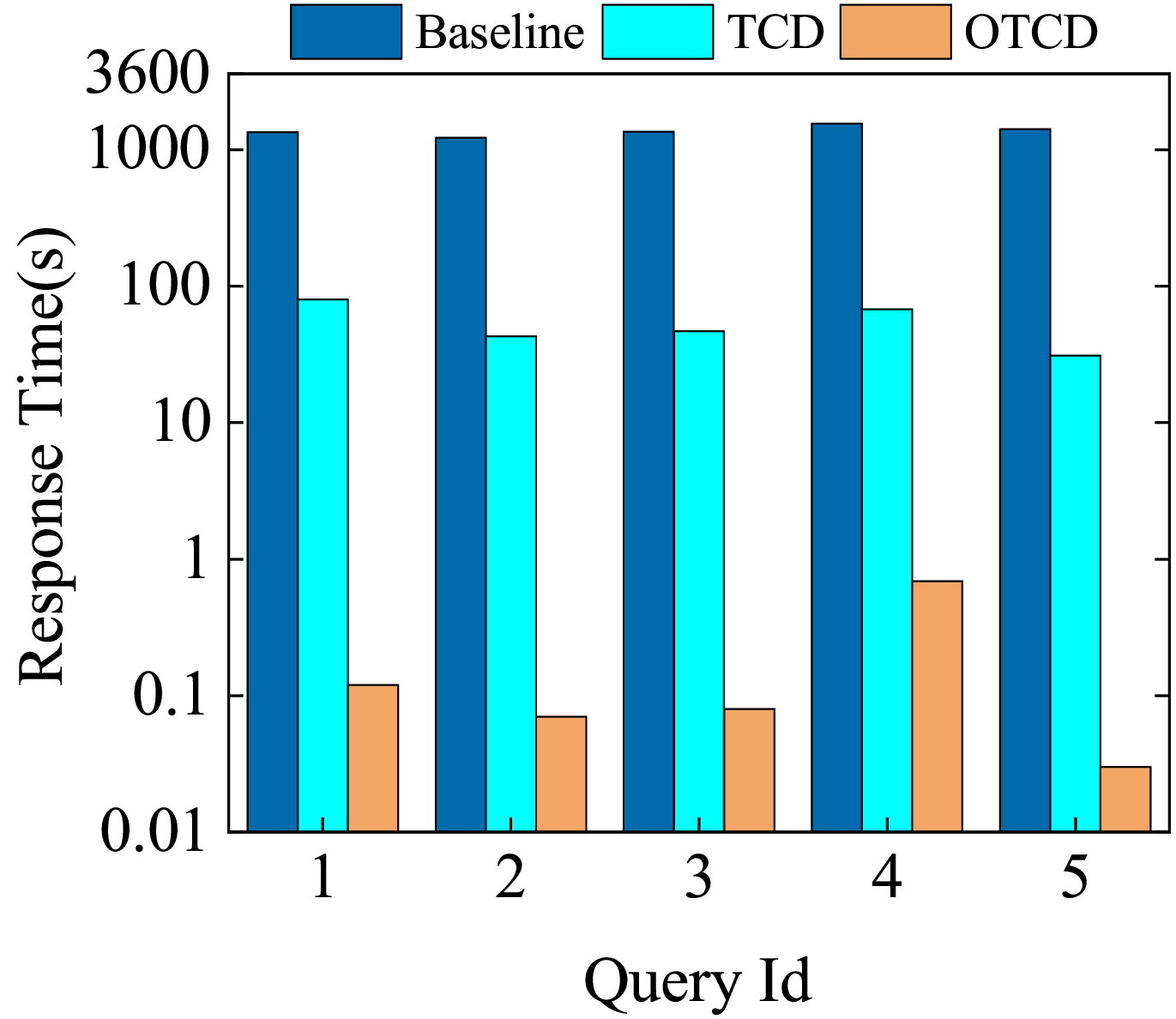}}
		\subfloat[email-Eu-core-temporal]{\label{subfig:email-Eu-core-temporal-colume}
			\includegraphics[width=0.5\linewidth]{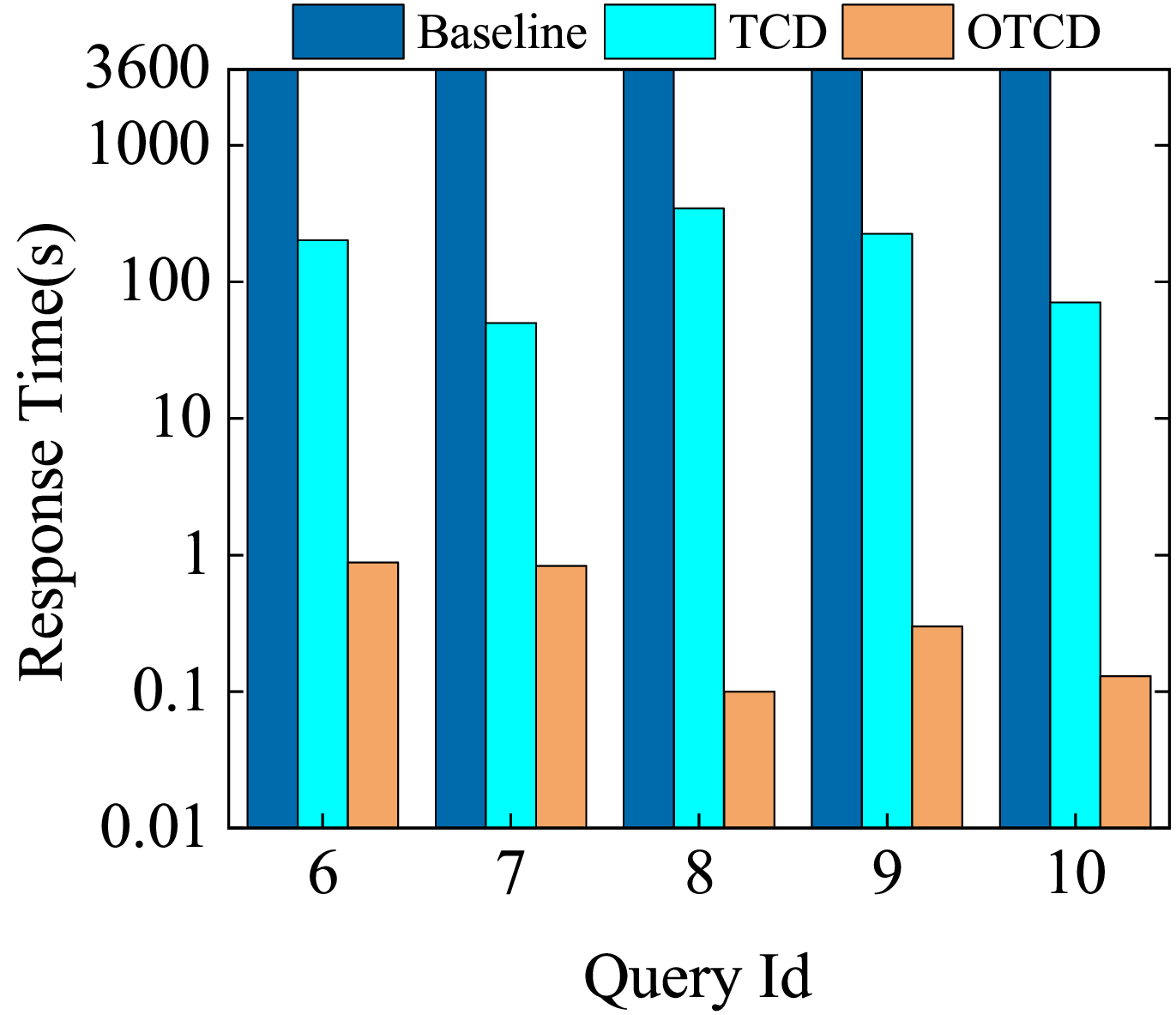}}
		\newline
		\subfloat[sx-mathoverflow]{\label{subfig:sx-mathoverflow-colume}
			\includegraphics[width=0.5\linewidth]{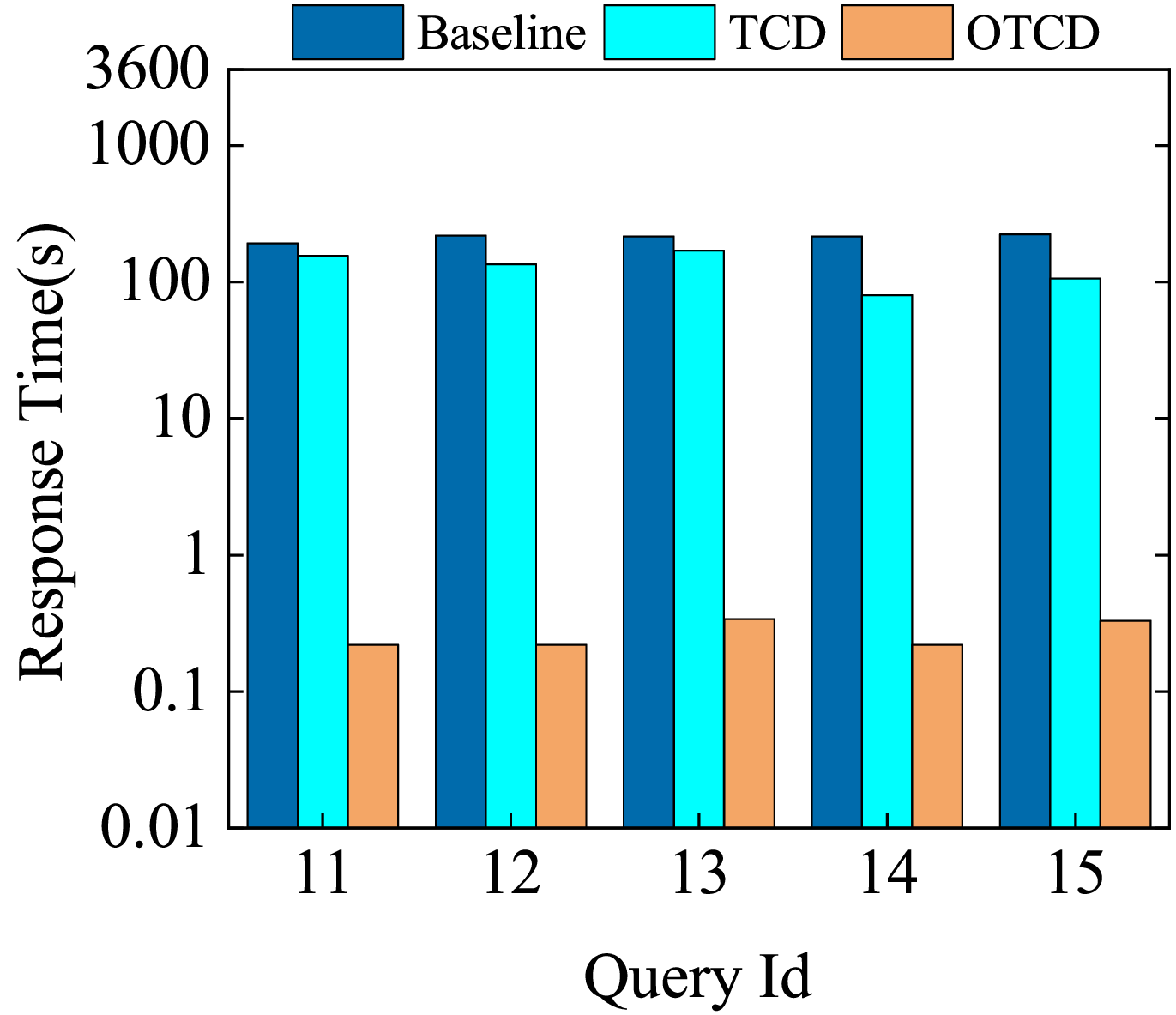}}
		\subfloat[sx-stackoverflow]{\label{subfig:sx-stackoverflow-colume}
			\includegraphics[width=0.5\linewidth]{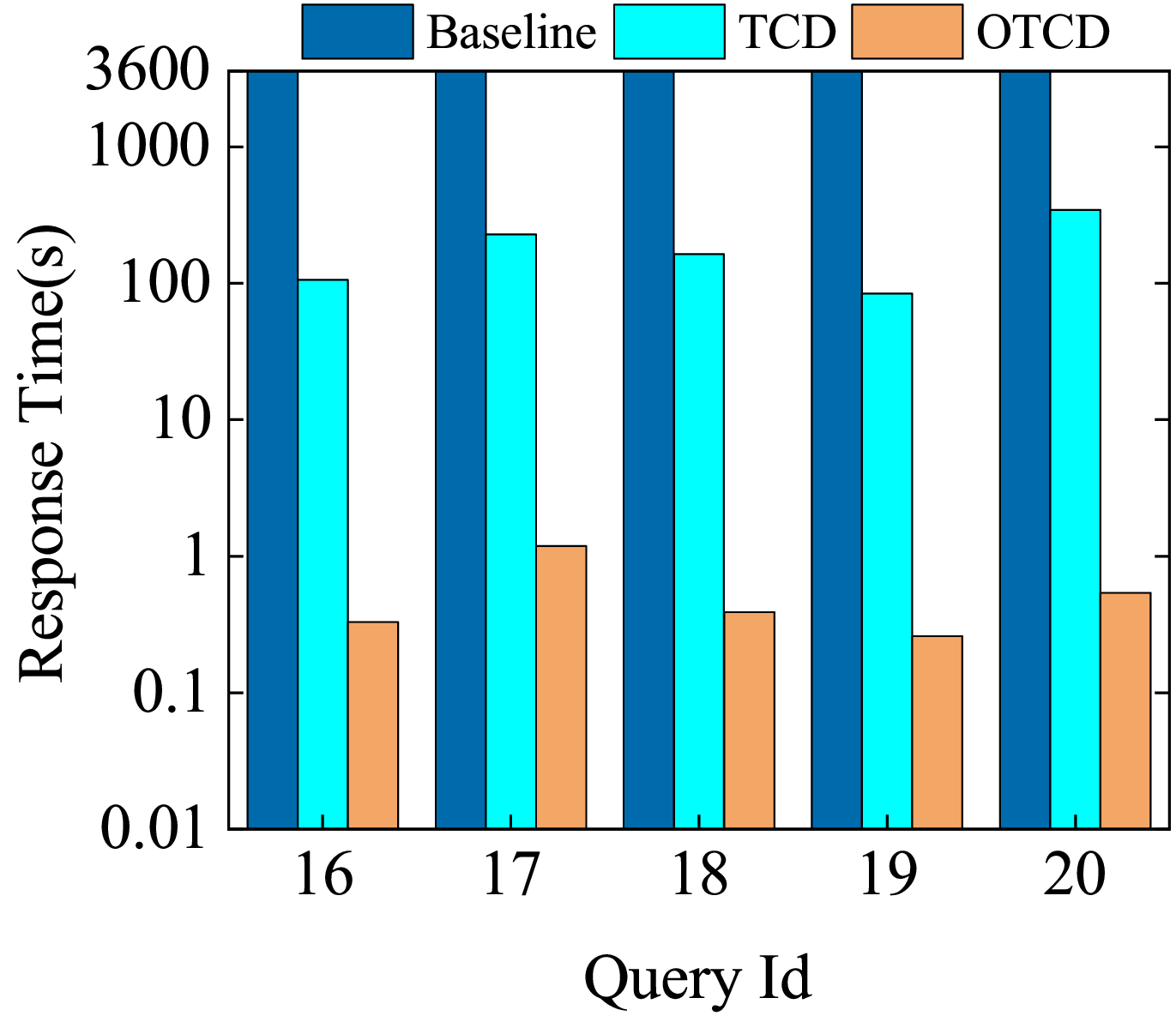}}
		
		\caption{The comparison of response time for selected queries on SNAP graphs.}\label{fig:colchart}
		
	\end{figure}

Figure~\ref{fig:colchart} compares the response time of Baseline (iPHC-Query), TCD and OTCD algorithms for each selected query respectively, which clearly demonstrates the efficiency of our algorithm. Firstly, TCD performs better than baseline for all twenty queries due to the physical efficiency of TEL, though they both decrementally or incrementally induce temporal $k$-cores. Specifically, TCD spends around 100 sec for each query. In contrast, baseline spends more than 1000 sec on CollegeMsg and even cannot finish within an hour on two other graphs, though it uses a precomputed index. Furthermore, OTCD runs two or three orders of magnitude faster than TCD, and only spends about 0.1-1 sec for each query, which verifies the effectiveness of our pruning method based on TTI. 
	
To compare the effect of three pruning rules in OTCD algorithm, Table~\ref{table:prune} lists their triggered times and the percentage of subintervals pruned by them for several queries respectively. PoR and PoU are triggered frequently because their conditions are more easily to be satisfied. However, PoR actually contributes pruned subintervals much less than the other two. Because it only prunes subintervals in the same row, and in contrast, PoU and PoL can prune an ``area'' of subintervals. Overall, the three pruning rules can achieve significant optimization effect together by enabling OTCD algorithm to skip more than 80 percents of subintervals.

\begin{figure}
		\centering
		\subfloat[Youtube]{\label{subfig:youtube}
			\includegraphics[width=0.5\linewidth]{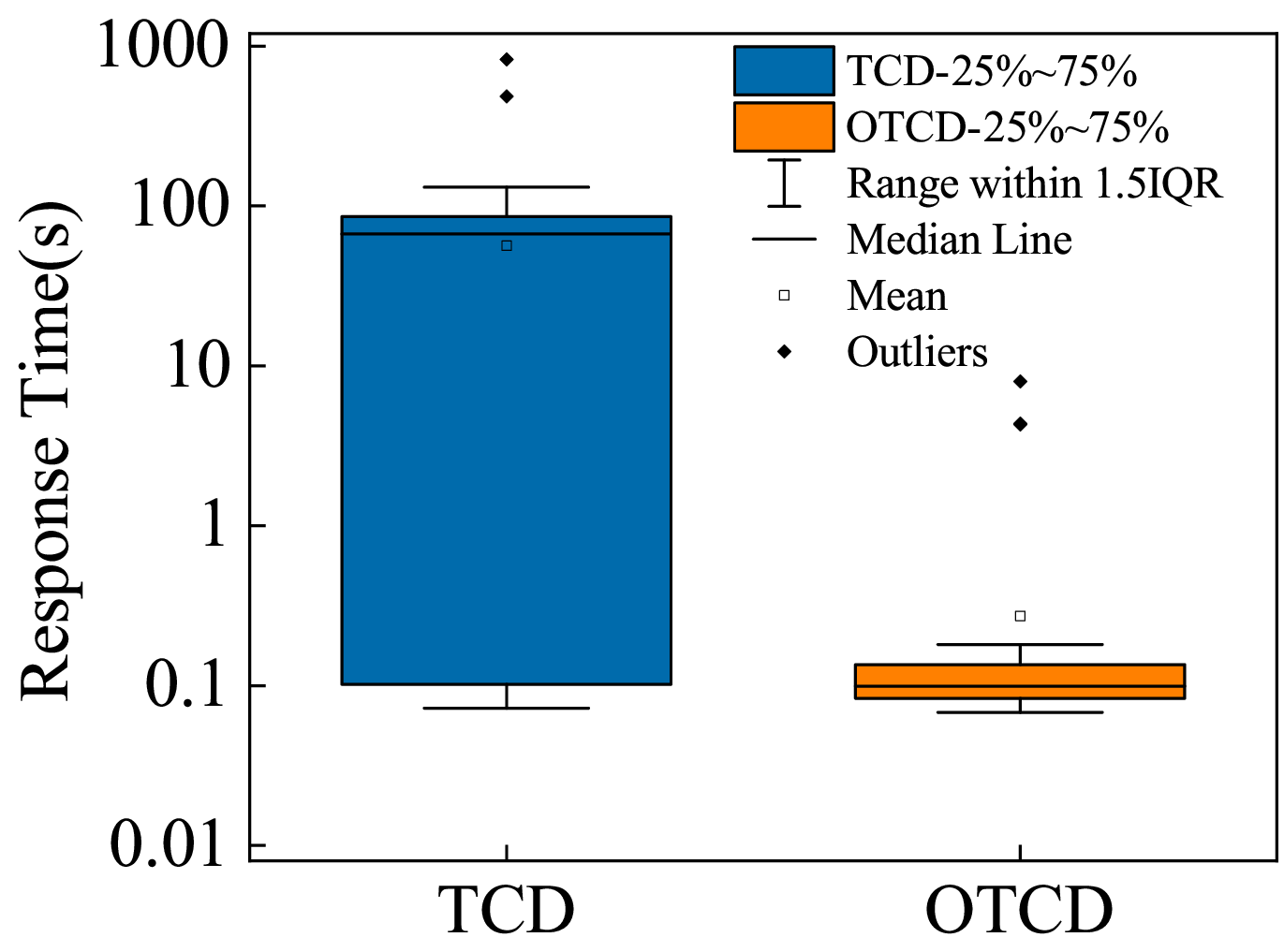}}
		\subfloat[Flickr]{\label{subfig:Flickr}
			\includegraphics[width=0.5\linewidth]{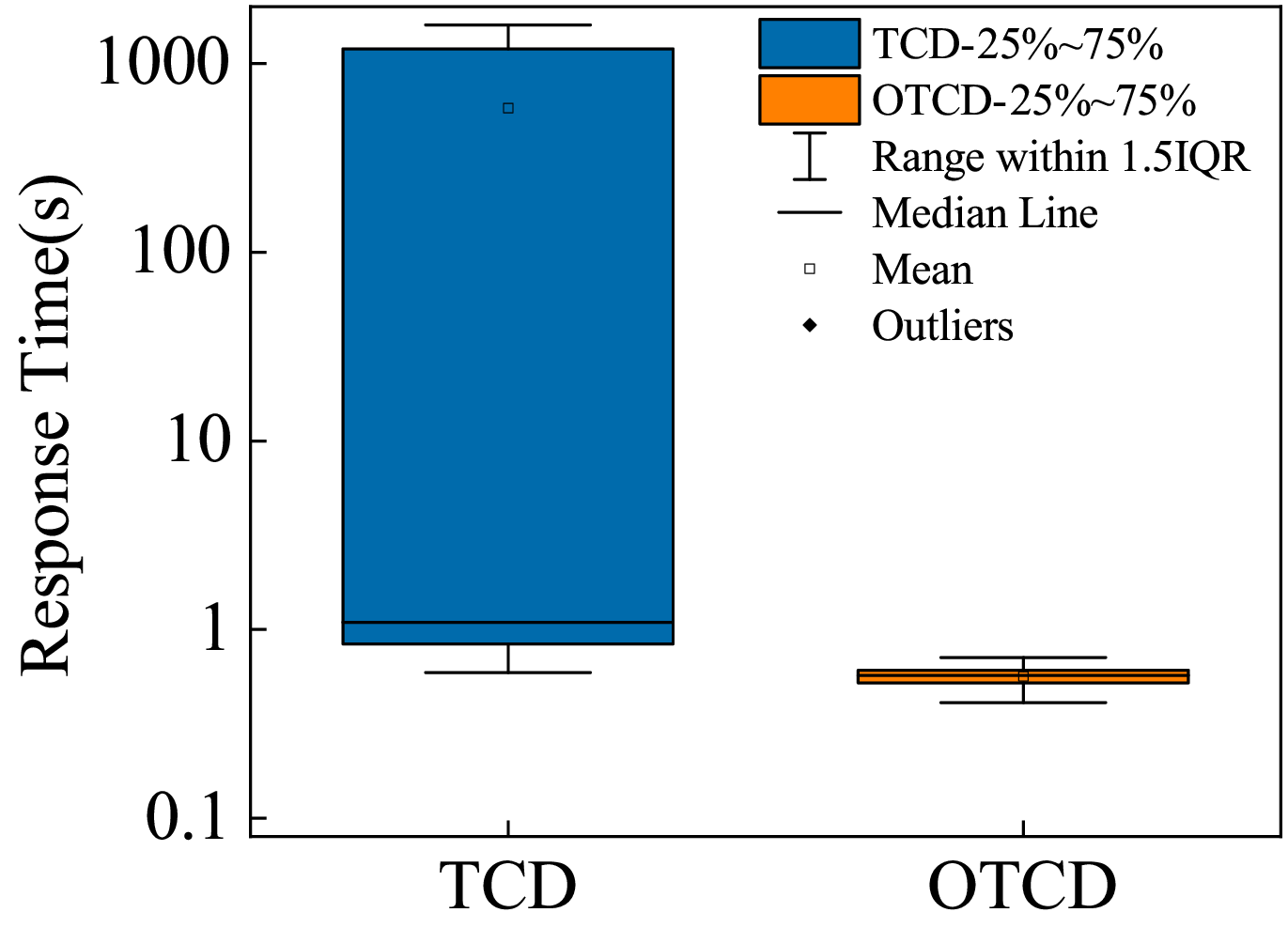}}
		\caption{The statistical distribution of response time for random queries on KONECT graphs.}\label{fig:box}
	\end{figure}

To evaluate the stability of our approach, we conduct statistical analysis of one hundred valid random queries on two new graphs, namely, Youtube and Flickr. We visualize the distribution of response time of TCD and OTCD algorithms for these random queries as boxplots, which are shown by Figure~\ref{fig:box}. The boxplots demonstrate that the response time of OTCD varies in a very limited range, which indicates that the OTCD indeed performs stable in practice. The outliers represent some queries that have exceptionally more results, which can be seen as a normal phenomenon in reality. They may reveal that many communities of the social networks are more active during the period.
	
\begin{table}
    \centering
    \caption{Memory consumption of (O)TCD algorithm.}
    \begin{tabular}{lc}
    \hline
    Dataset & Process Memory (GB)\\
    \hline
    CollegeMsg & 0.02\\
    sx-mathoverflow & 0.06\\
    Youtube & 1.7\\
    DBLP & 3.1\\
    Flickr & 3.5\\
    sx-stackoverflow & 6.5\\
    \hline
    \end{tabular}
	\label{tab:procmem}
\end{table}

Moreover, Table~\ref{tab:procmem} reports the process memory consumption for different datasets, which depends on the size of TEL mostly. We can observe that, 1) for the widely-used graphs like Youtube, DBLP, Flickr and stackoverflow, several gigabytes of memory are needed for storing TEL, which is acceptable for the ordinary hardware; and 2) for the very large graphs with billions of edges, the size of TEL is hundreds of gigabytes approximately, which would require the distributed memory cluster like Spark.
	
To verify the scalability of our method with respect to the query parameters, we test the three algorithms with varing minimum degree $k$ and time span (namely, $Te-Ts$) respectively.

\begin{figure}
		\centering
		\subfloat[CollegeMsg]{\label{subfig:collegemsg-k}
			\includegraphics[width=0.333\linewidth]{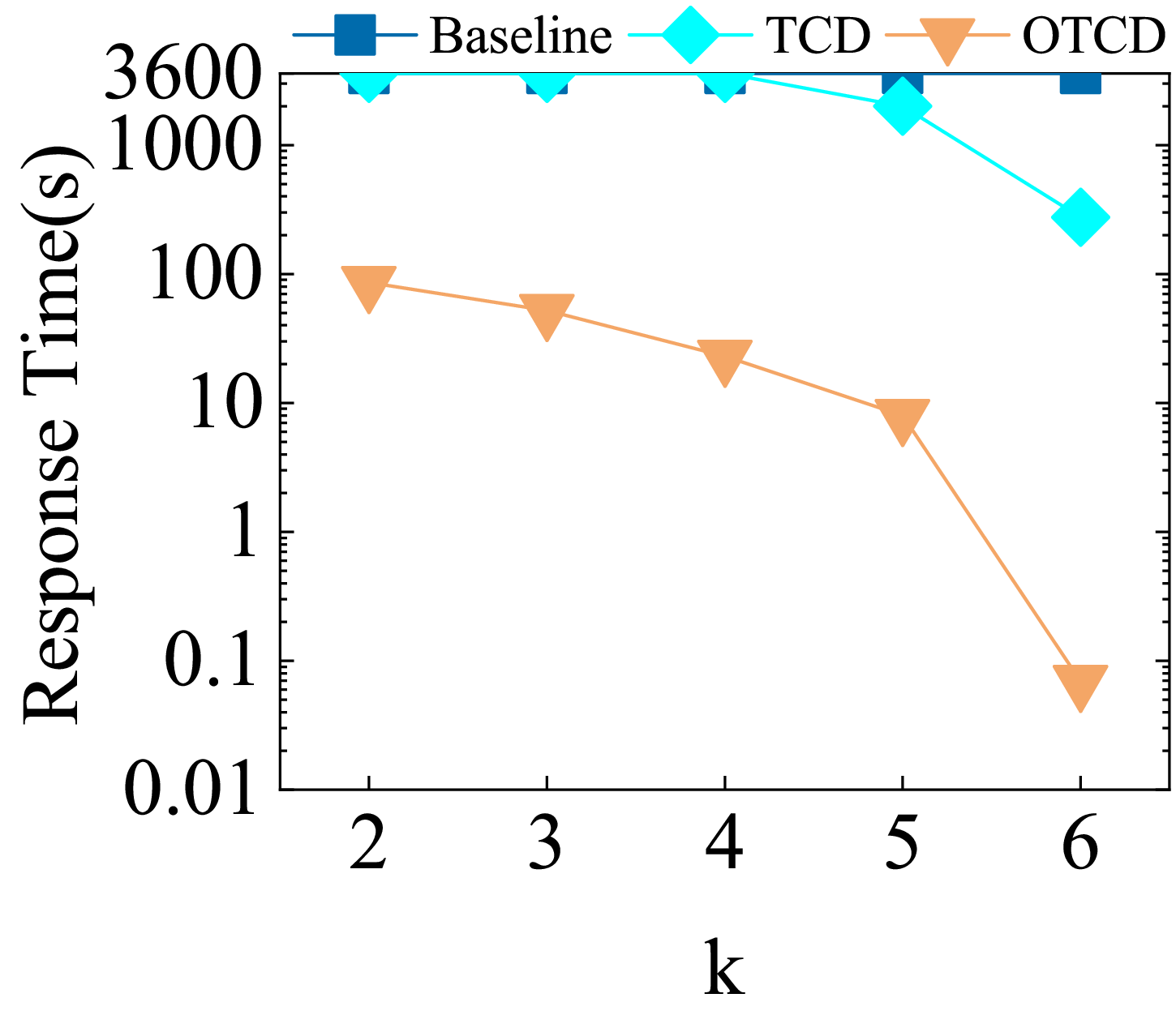}}
		\subfloat[sx-mathoverflow]{\label{subfig:sx-mathoverflow-k}
			\includegraphics[width=0.333\linewidth]{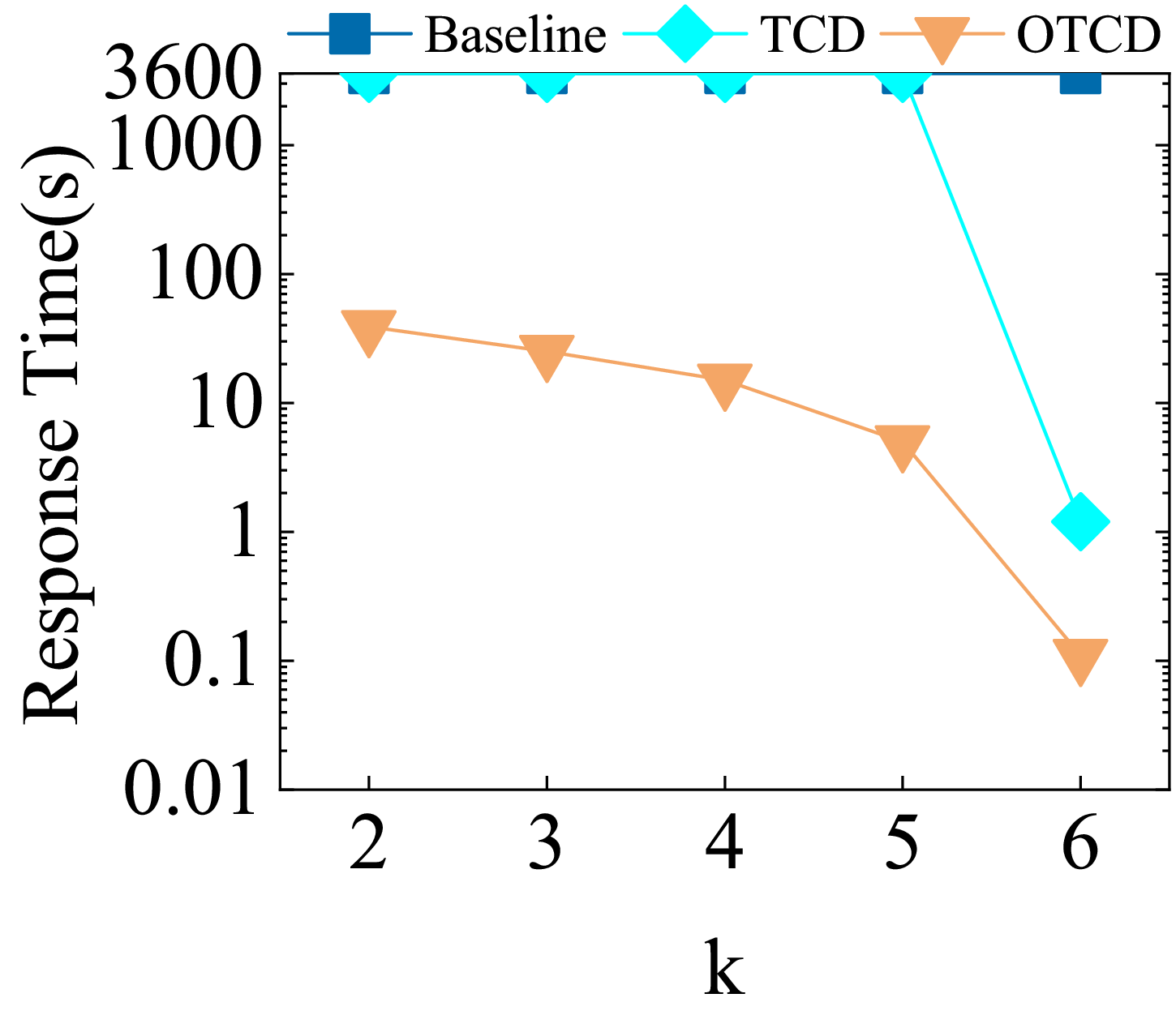}}
		\subfloat[sx-stackoverflow]{\label{subfig:sx-stackoverflow-k}
			\includegraphics[width=0.333\linewidth]{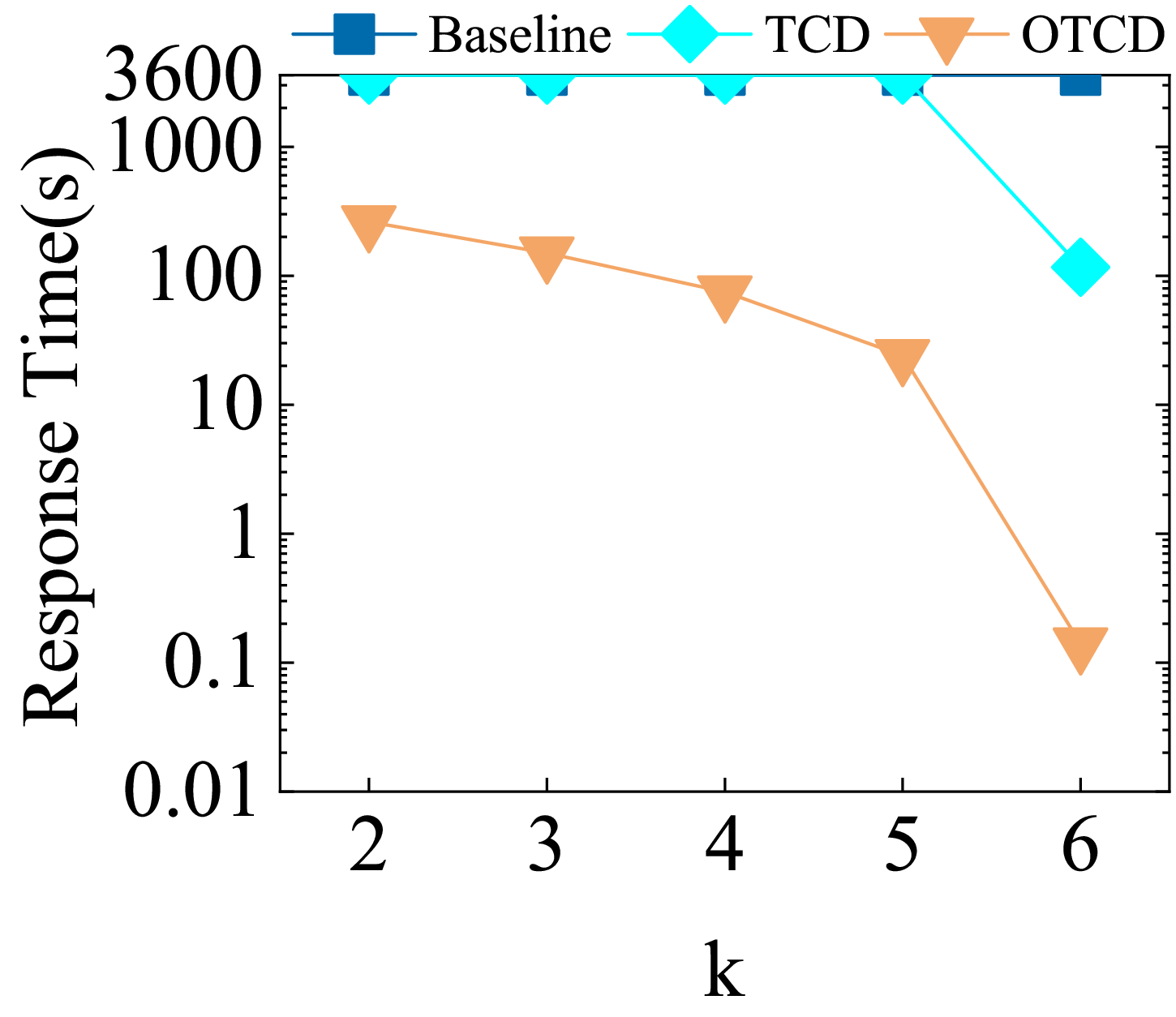}}
		\caption{Trend of response time under a range of $k$.}\label{fig:kchart}
	\end{figure}

\begin{figure}
		\centering
		\subfloat[CollegeMsg]{\label{subfig:collegemsg-kn}
			\includegraphics[width=0.333\linewidth]{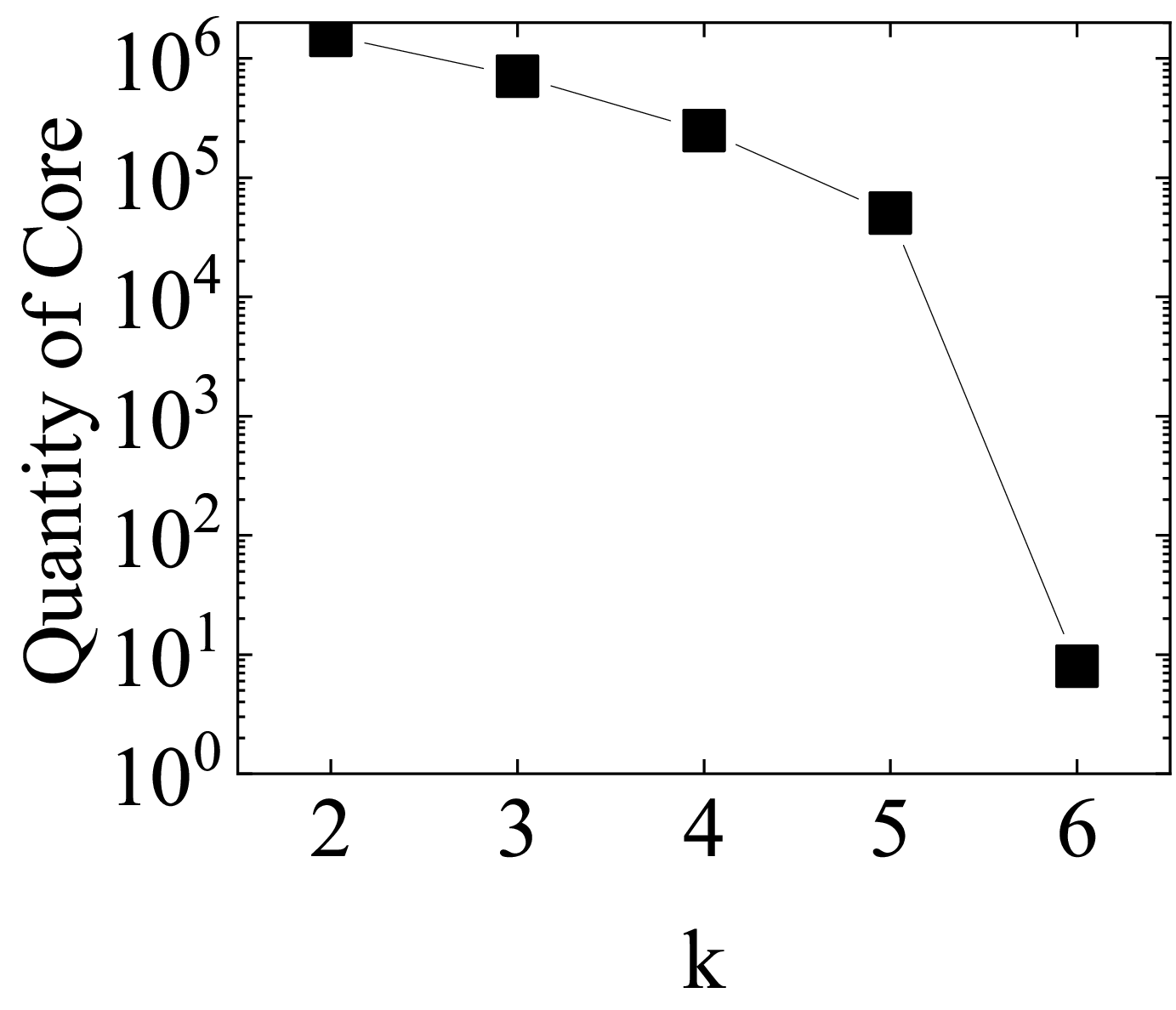}}
		\subfloat[sx-mathoverflow]{\label{subfig:sx-mathoverflow-kn}
			\includegraphics[width=0.333\linewidth]{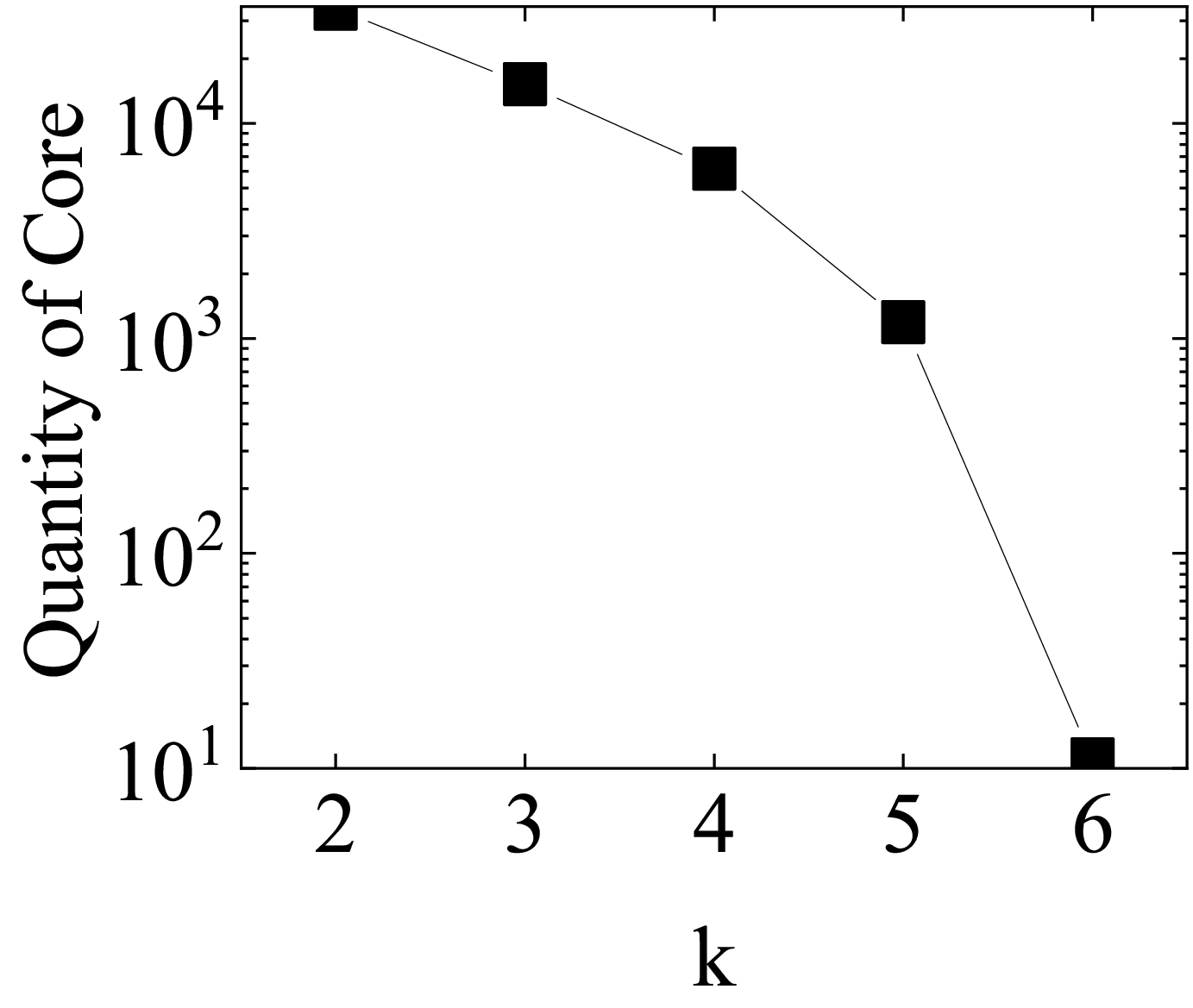}}
		\subfloat[sx-stackoverflow]{\label{subfig:sx-stackoverflow-kn}
			\includegraphics[width=0.333\linewidth]{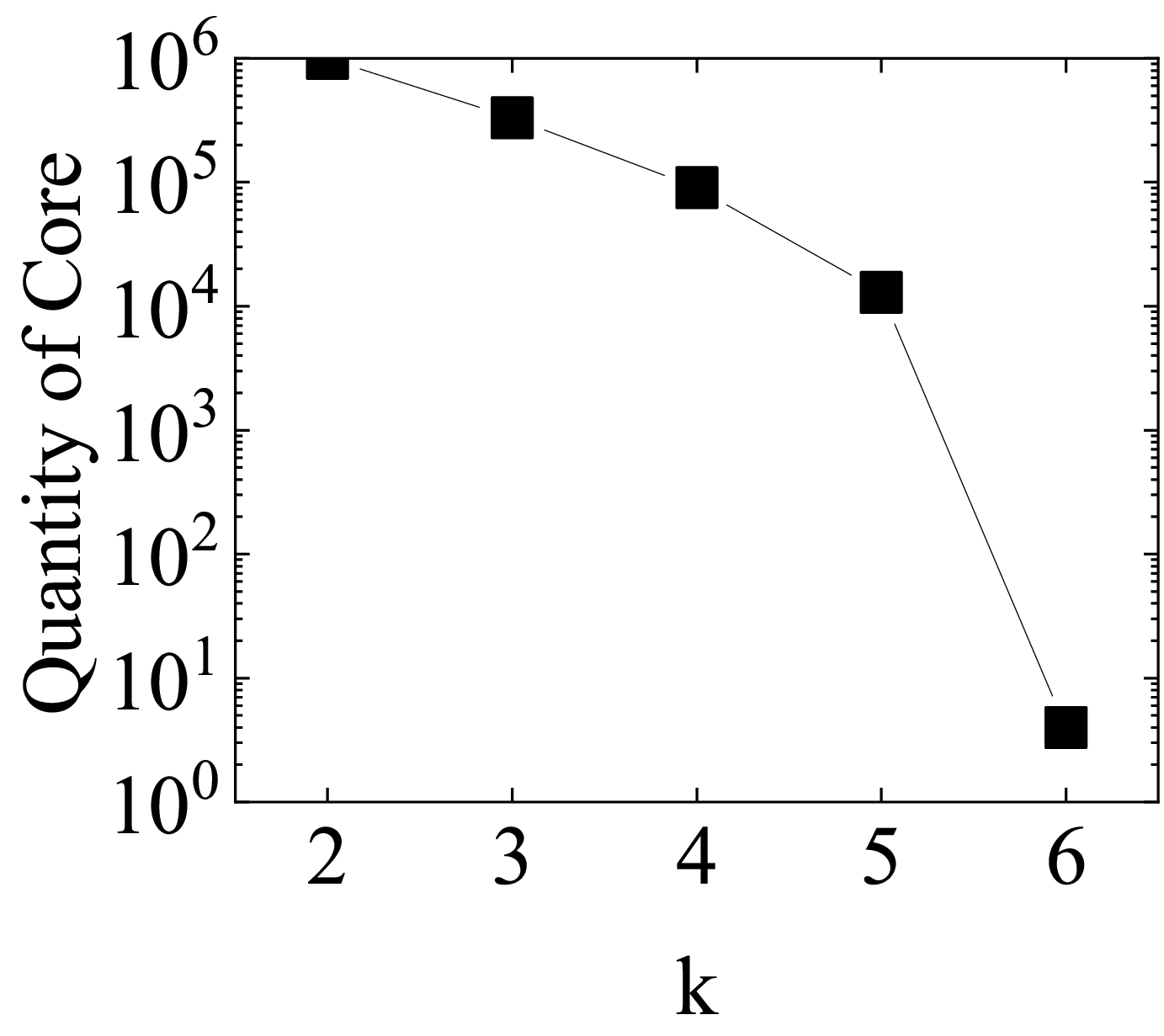}}
		\caption{Trend of amount of distinct temporal $k$-cores under a range of $k$.}\label{fig:kchart2}
	\end{figure}

\begin{figure}
		\centering
		\subfloat[CollegeMsg]{\label{subfig:collegemsg-kscc}
			\includegraphics[width=0.333\linewidth]{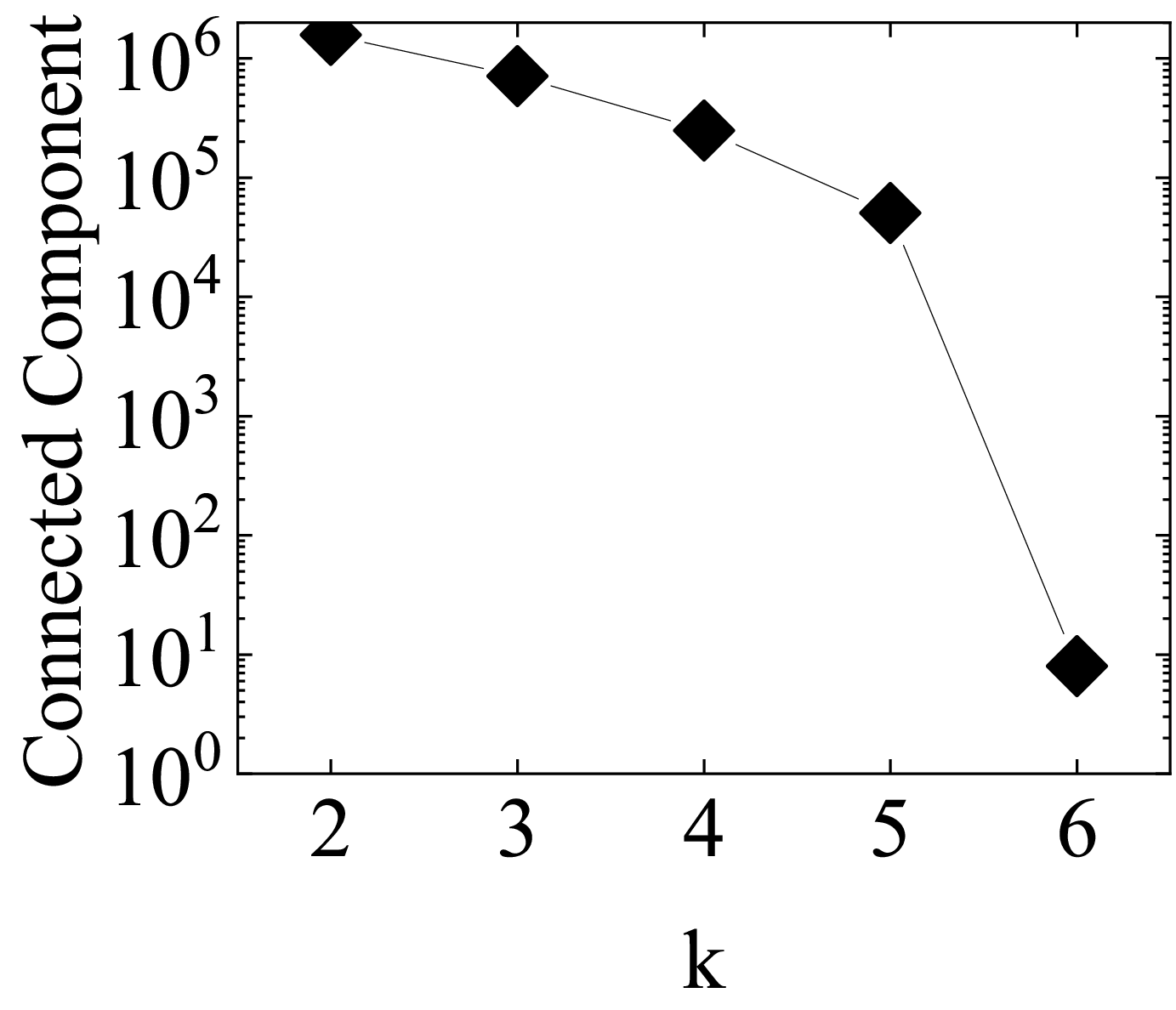}}
		\subfloat[sx-mathoverflow]{\label{subfig:sx-mathoverflow-kscc}
			\includegraphics[width=0.333\linewidth]{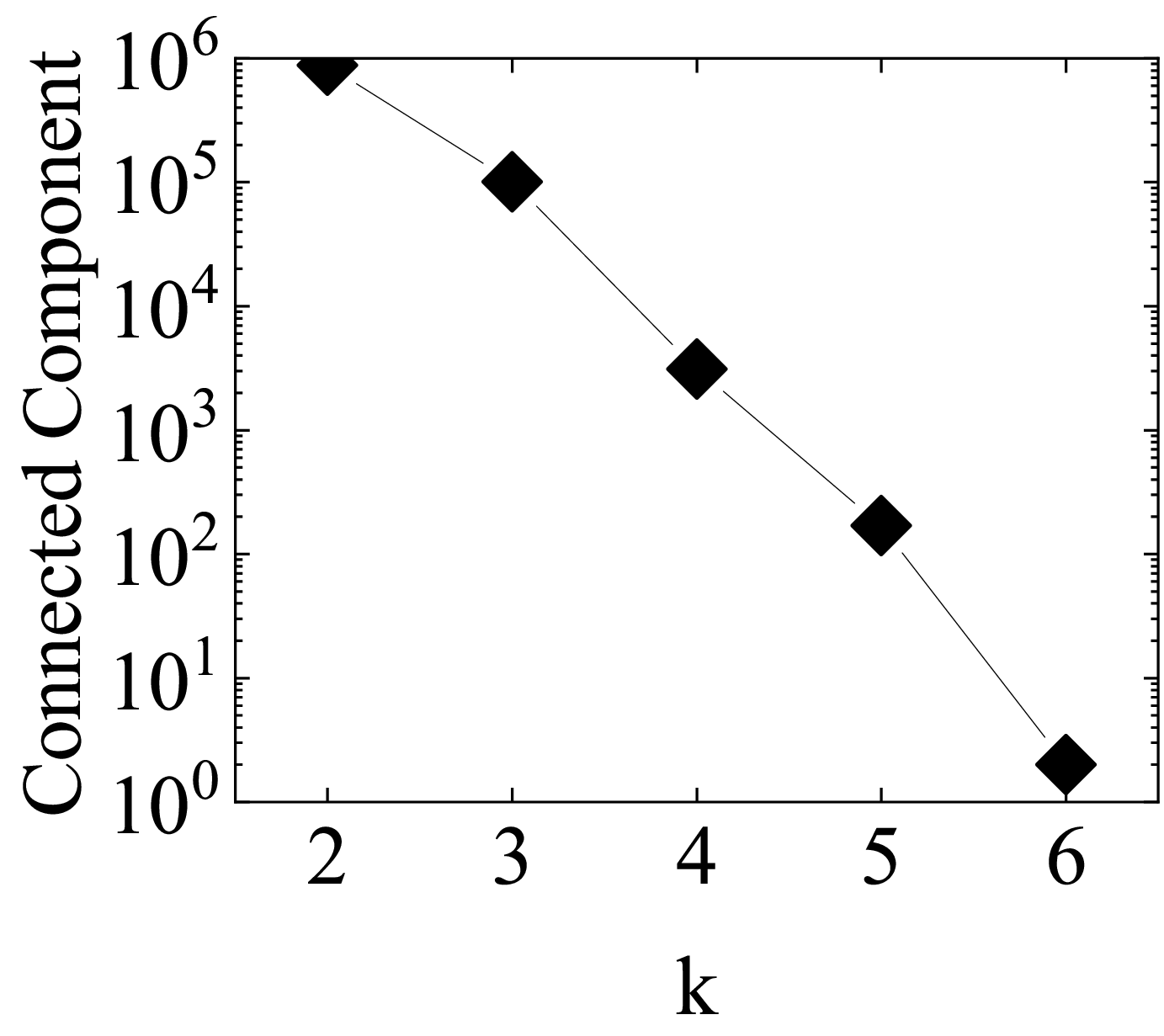}}
		\subfloat[sx-stackoverflow]{\label{subfig:sx-stackoverflow-kscc}
			\includegraphics[width=0.333\linewidth]{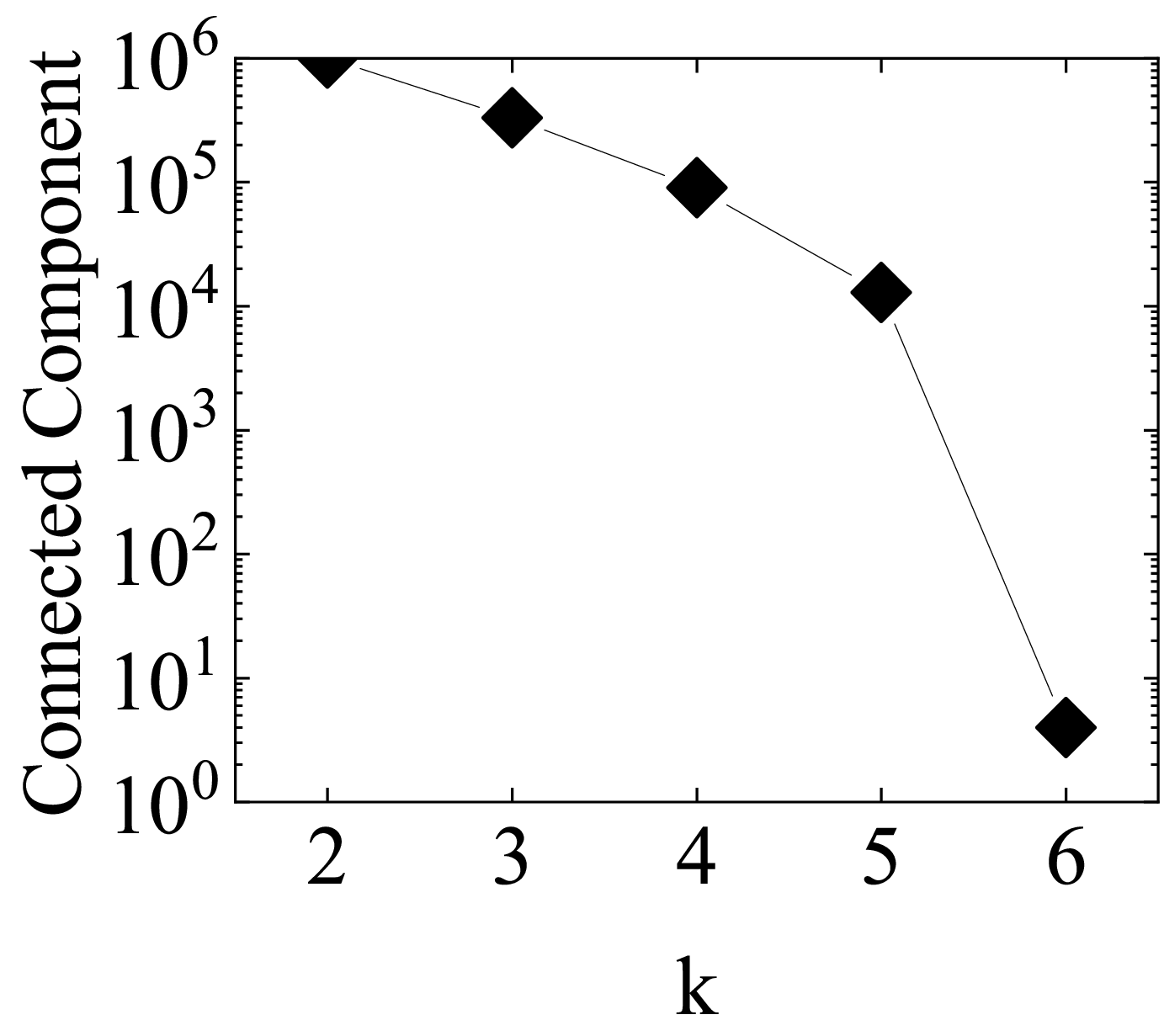}}
		\caption{Trend of amount of connected components in temporal $k$-cores under a range of $k$.}\label{fig:kchart3}
	\end{figure}

\textbf{Impact of $k$.} We select a typical query with span fixed and $k$ ranging from 2 to 6 for different graphs. The response time of tested algorithms are presented in Figure~\ref{fig:kchart}, from which we have an important observation against common sense. That is, different from core decomposition on non-temporal graphs, when the value of $k$ increases, the response time of TCD and OTCD algorithms decreases gradually. For OTCD, the behind rationale is clear, namely, its time cost is only bounded by the scale of results, which decreases sharply with the increase of $k$. To support the claim, Figure~\ref{fig:kchart2} and Figure~\ref{fig:kchart3} show the trend of the amount of result cores and connected components in the result cores changing with $k$. Intuitively, a greater value of k means a stricter constraint and thereby filters out some less cohesive cores. We can see the trend of runtime decrease for OTCD in Figure~\ref{fig:kchart} is almost the same as the trend of core amount decrease in Figure~\ref{fig:kchart2}, which also confirms the scalability of OTCD algorithm. For TCD, the behind rationale is complicated, since it enumerates all subintervals and each single decomposition is more costly with a greater value of $k$. It is just like peeling an onion layer by layer, which has less layers with a greater value of $k$, so that the maintenance between layers become less.

\begin{figure}
		\subfloat[CollegeMsg]{\label{subfig:CollegeMsg-Span}
			\includegraphics[width=0.333\linewidth]{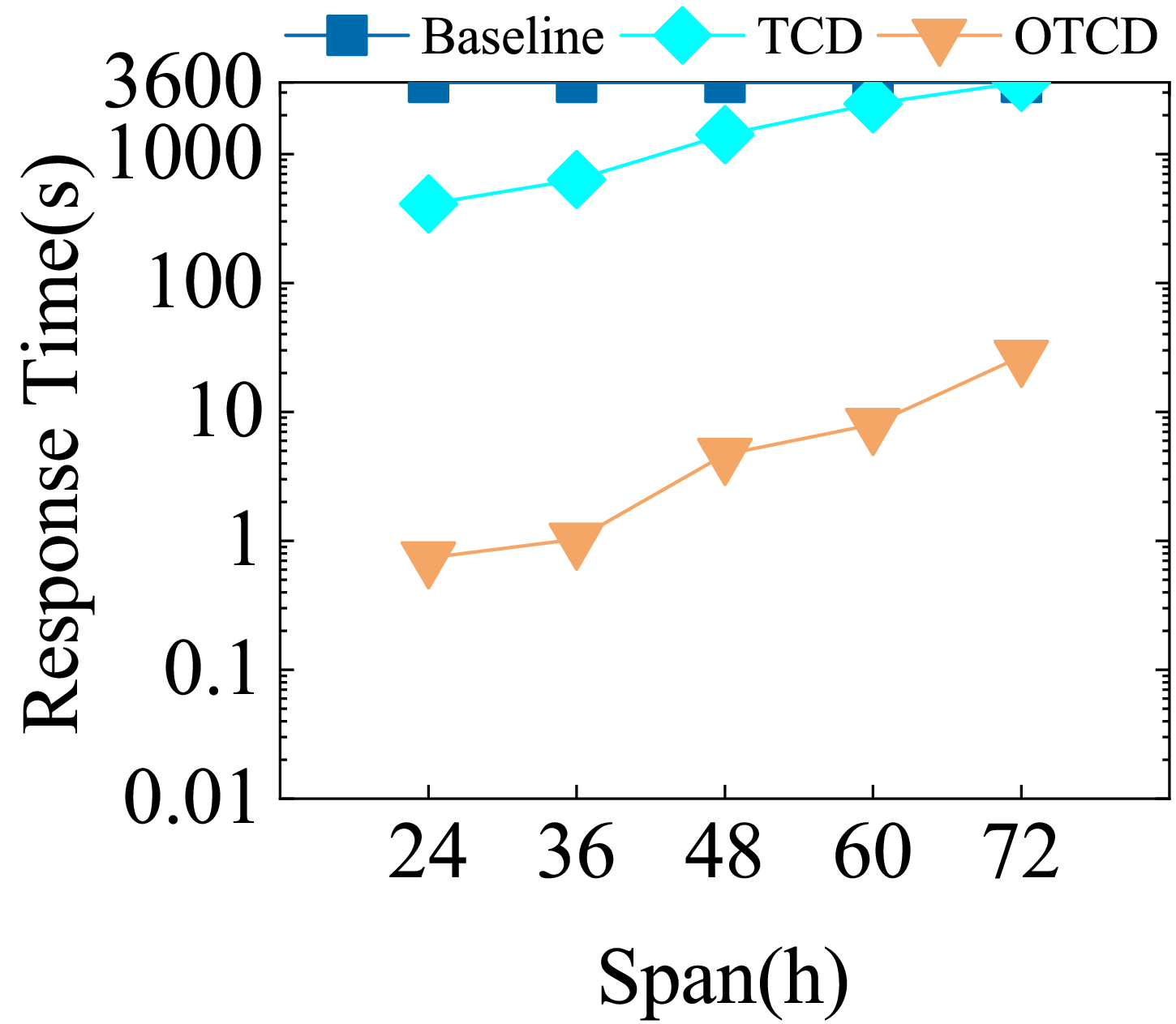}}
		\subfloat[sx-mathoverflow]{\label{subfig:sx-mathoverflow-Span}
			\includegraphics[width=0.333\linewidth]{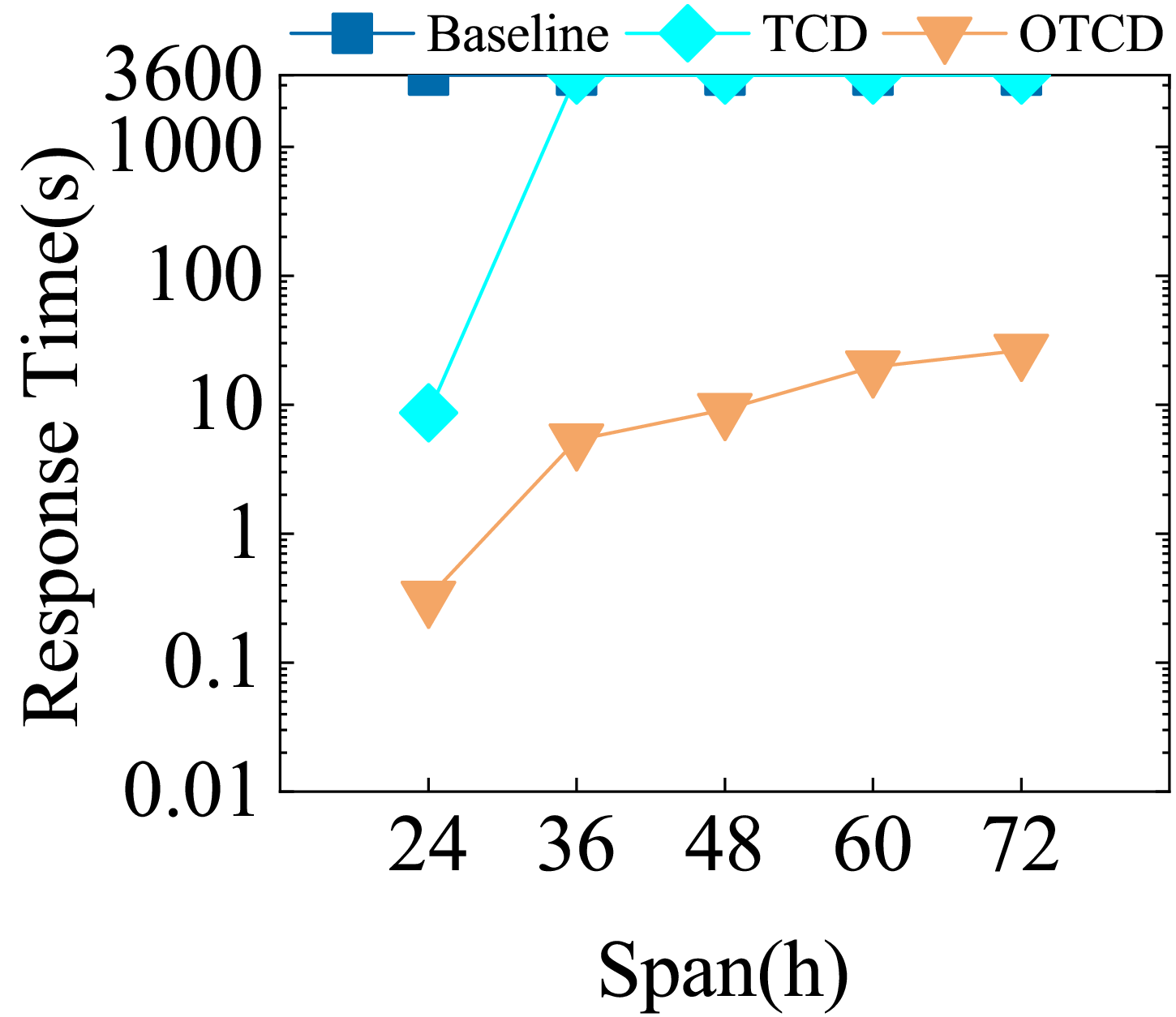}}
		\subfloat[sx-stackoverflow]{\label{subfig:sx-stackoverflow-Span}
			\includegraphics[width=0.333\linewidth]{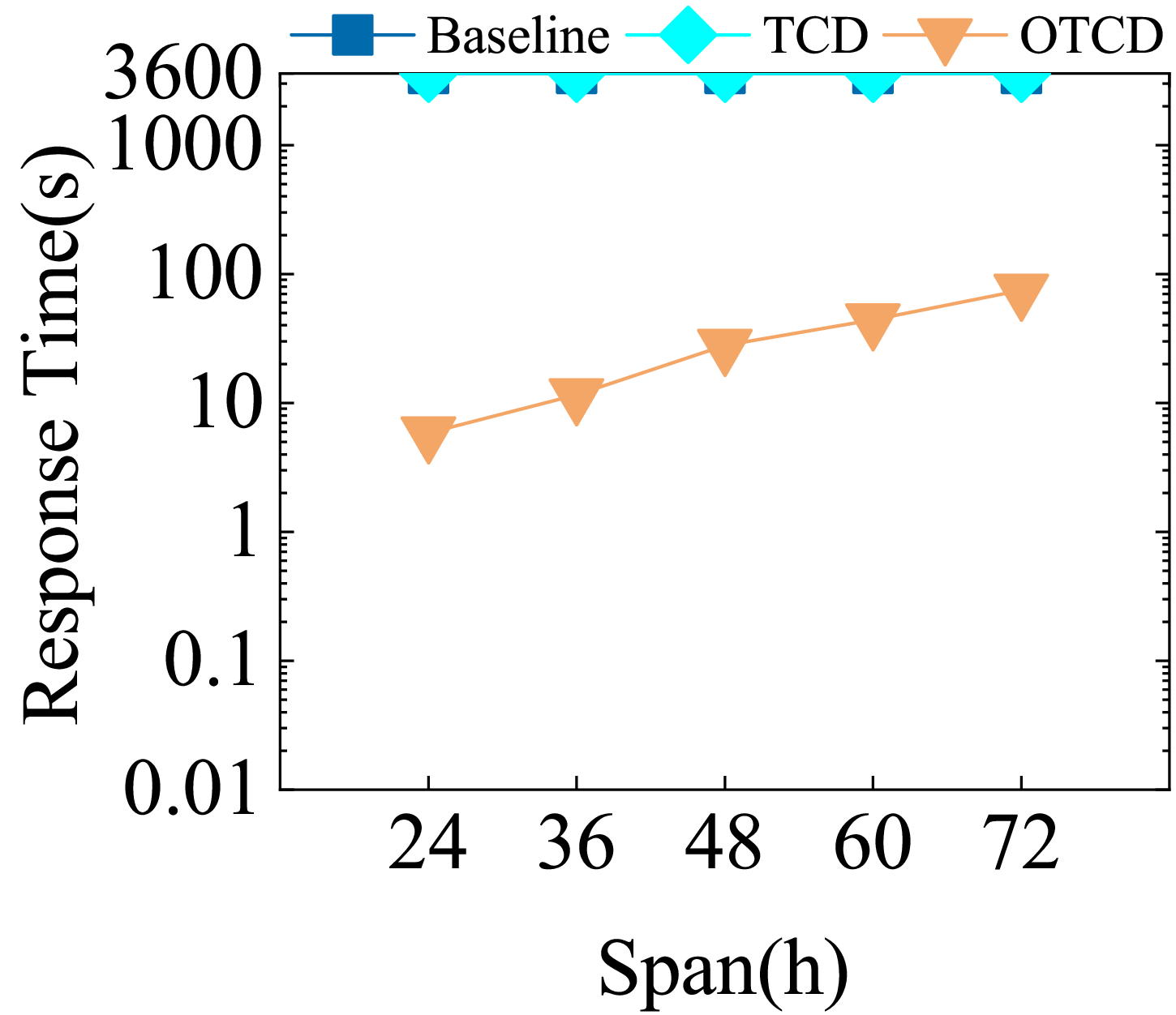}}
		\caption{Trend of response time under a range of span.}\label{fig:spanchart}	
	\end{figure}

\textbf{Impact of span.} Similarly to the test of $k$, we also evaluate the scalability of different algorithms when the query time span increases. The results are presented in Figure~\ref{fig:spanchart}. Although the number of subintervals increases quadratically, the response time of OTCD still increases moderately following the scale of query results. In contrast, TCD runs dramatically slower when the query time span becomes longer.

The above results demonstrate that the efficiency of OTCD is not sensitive to the change of query parameters, so that it is scalable in terms of query time interval.

Lastly, for a large graph with a long time span like Youtube, we test OTCD algorithm by querying temporal 10-cores over the whole time span. The result is, to find all 19,146 temporal 10-cores within 226 days, the OTCD algorithm spent about 55 minutes, which is acceptable for such a ``full graph scan'' task.
	
\subsection{Effectiveness}

The effectiveness of TCQ is two-fold. Firstly, by given a flexible time interval, we can find many temporal $k$-cores of different subintervals, each of which represents a community emerged in a specific period. Consider the above test on Youtube. Although it is not feasible to exhibit all 19,146 cores, Figure~\ref{fig:scale} shows their distribution by time span. The number of cores generally decreases with the increase of time span, which makes sense because there are always a lot of small communities emerged during short periods and then they will interact with each other and be merged to larger communities within a longer time span. 

Secondly, we can continue to filter and analyse the result cores to gain insights. For example, we record the date in GMT time for nine of the result cores with a time span less than one day in Youtube, and try to figure out if they emerged for some special reasons. Table~\ref{table:tkc0span} lists the date and size of the nine cores. We can see that there is a large core emerged on Dec 10, 2006, which means more than 40,000 accounts had nearly one million interactions with each other in just a day. That is definitely caused by a special event. While, most of the rest cores emerged during summer vacation, which may mean people have more interactions on Youtube in the period.

\noindent\begin{minipage}[p]{0.49\linewidth}
		\includegraphics[width=\textwidth]{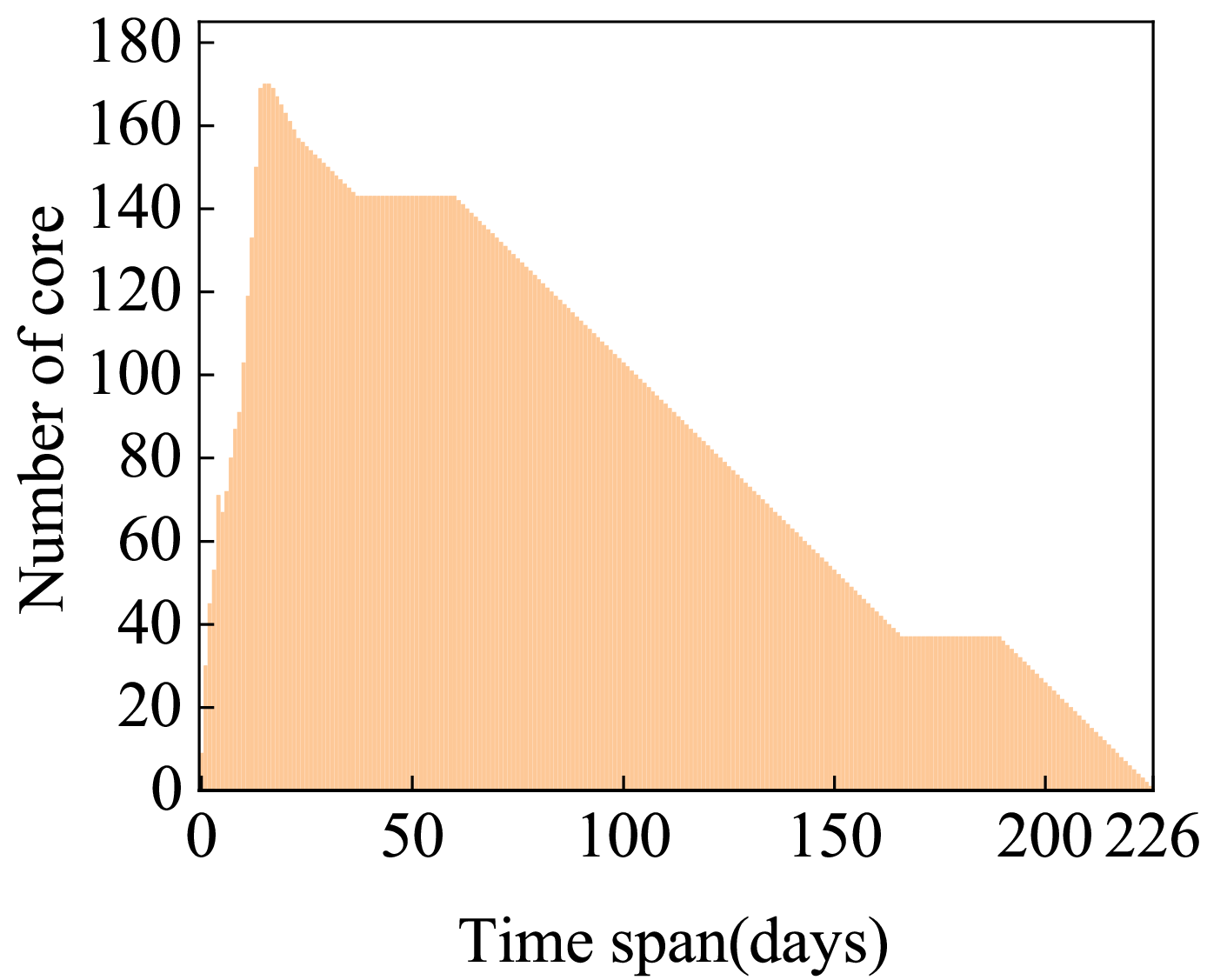}
		\captionof{figure}{Distribution of all temporal 10-cores in Youtube by time span.}
		\label{fig:scale}
	\end{minipage}
	\hspace{0.01\textwidth}
	\begin{minipage}[p]{0.49\linewidth}
		\captionof{table}{The date and size of nine temporal 10-cores emerged within one day in Youtube.}
		\begin{tabular}{lll}
			\hline
			Date & |$\mathcal{V}|$ & |$\mathcal{E}$|\\
			\hline
			Dec 10 2006 & 46499 & 885128\\
			Feb 08 2007 & 1268 & 12054\\
			Mar 25 2007 & 21 & 139\\
			Jun 15 2007 & 98 & 713\\
			Jun 18 2007 & 20 & 100\\
			Jun 20 2007 & 124 & 1012\\
			Jun 30 2007 & 21 & 110\\
			Jul 02 2007 & 21 & 110\\
			Jul 06 2007 & 12 & 66\\
			\hline
		\end{tabular}
		\label{table:tkc0span}
\end{minipage}

\subsection{Case Study}

\begin{figure}[t]
		\centering
		\includegraphics[width=\linewidth]{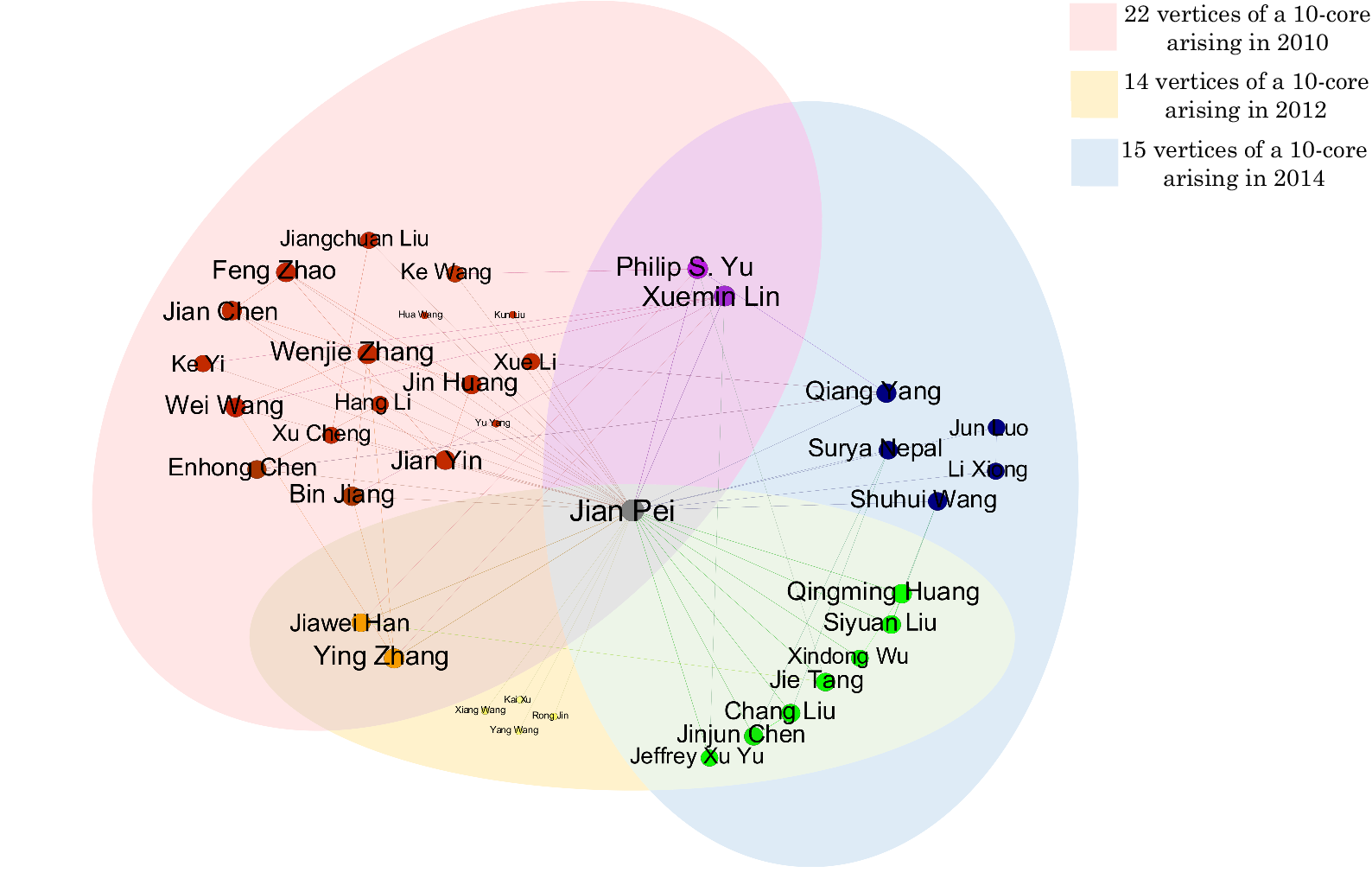}
		\caption{Case Study in DBLP coauthorship network.}\label{fig:cstudydblp}
	\end{figure}

For case study, we employ OTCD algorithm to query temporal 10-cores on DBLP. The query interval is set as 2010 to 2018, which spans over 8 years. By statistics, there exist 43 temporal 10-cores during that period, with 39 of them containing the author Jian Pei, for whom we further build an ego network from three selected cores in defferent years. Figure~\ref{fig:cstudydblp} shows the ego network. The authors in the three cores emerged in 2010, 2012 and 2014 are shaded by red, yellow and blue respectively. By observing the evolution of ego network over years, we can infer the change of author's research interests or affiliations.

\begin{figure}[t]
		\centering
		\includegraphics[width=0.8\linewidth]{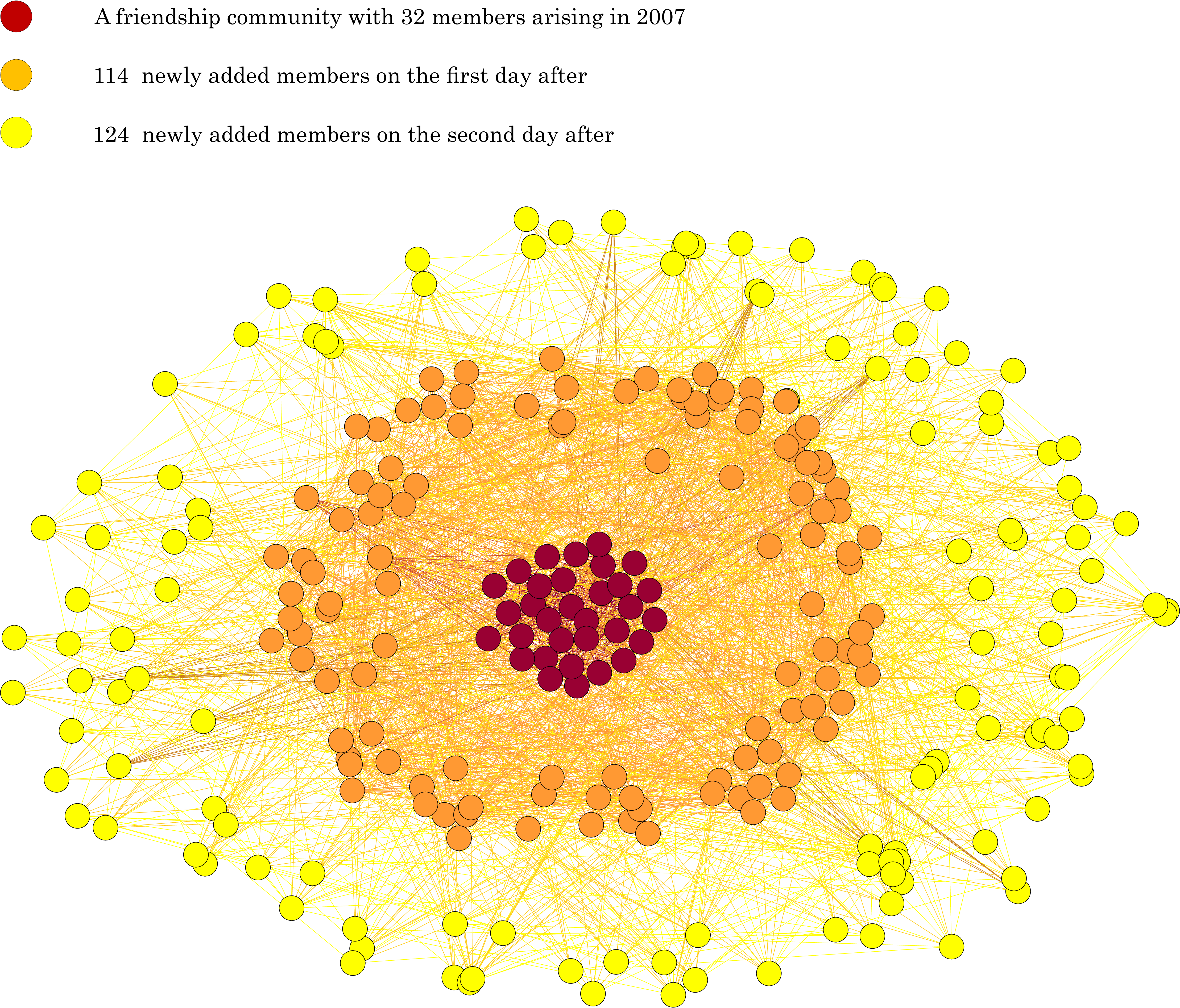}
		\caption{Case Study in Youtube friendship network.}\label{fig:cstudyyoutube}
	\end{figure}

To further demonstrate the potential of TCQ, we also employ TCQ to find temporal $k$-cores that expand quickly over time. This topic has been addressed in \cite{chu2019online}. Since OTCD returns all distinct cores efficiently, we can conveniently achieve the goal by identifying the cores contained by other larger cores within a few of days from the results. Figure~\ref{fig:cstudyyoutube} shows such a bursting community on Youtube friendship network. The 32 central vertices colored in red comprise an initial temporal 10-core within two days. This core is contained by another core about four times larger, while the TTI of the larger core only expands by one day. The new vertices in the larger core are colored in orange. Then, the new vertices colored in yellow join them to comprise a twice larger new core in the next day. Clearly, these three temporal 10-cores together represent a community that grows remarkably fast. In the real world, with more concrete information of graphs, such usages of TCQ will facilitate applications like recommendation, disease control, etc.

	\begin{figure}[t]
		\centering
		\includegraphics[width=0.8\linewidth]{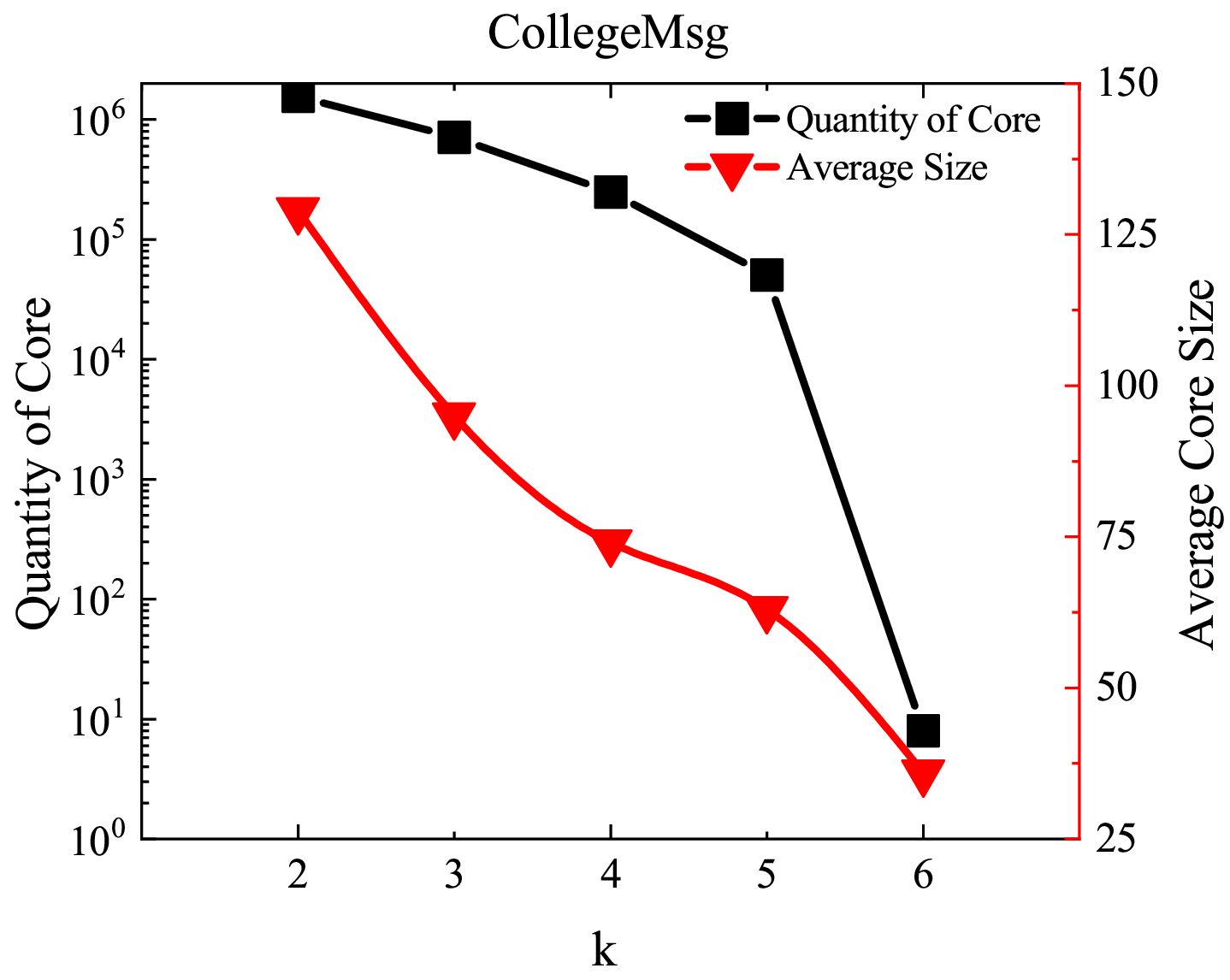}
		\caption{A statistical chart for selecting the value of $k$.}\label{fig:selectk}
	\end{figure}

\subsection{Discussion on the value of $k$}

TCQ achieves relaxing the constraint on query time interval when composing $k$-core queries on temporal graphs. However, the value of $k$ is still needed as an input parameter. We give a simple and rational criteria here for selecting the proper $k$ value on different graphs, though many potential factors have different impacts on the selection. The criteria is based on two intuitive facts. Firstly, the number of distinct temporal $k$-cores over a given time interval will decrease with the increase of $k$. Secondly, the size of returned temporal $k$-cores will shrink with the increase of $k$. Normally, we expect the result cores to be concise and non-overlapping, especially when detecting the suspicious communities that are inherently small and isolated, thereby preferring a greater value of $k$. However, the number of result cores also matters, which requires the value of $k$ not being too great, otherwise there could be too few results. Therefore, the selection of $k$ should take both size and number of result cores into account, just like the trade-off between precision and recall.

For example, with $k$ ranging from 2 until 6, Figure~\ref{fig:selectk} shows the falling curves of both number and average size of result cores over a specific time interval on CollegeMsg. We can observe that, setting $k = 5$ should be a good choice, since the core size has declined to a relatively small level while the number of results is still fairly sufficient.

\section{Related Work}

Recently, a variety of $k$-core query problems have been studied on temporal graphs, which involve different temporal objectives or constraints in addition to cohesiveness. The most relevant work to ours is historical $k$-core query~\cite{yu2021querying}, which gives a fixed time interval as query condition. In contrast, our temporal $k$-core query flexibly find cores of all subintervals. Moreover, Galimberti et al~\cite{galimberti2018mining} proposed the span-core query, which also gives a time interval as query condition. However, the span-core requires all edges to appear in every moment within the query interval, which is too strict in practice. Actually, historical $k$-core relaxes span-core, and temporal $k$-core further relaxes historical $k$-core. 

Besides, there are the following related work. Wu et al~\cite{wu2015core} proposed $(k,h)$-core and studied its decomposition algorithm, which gives an additional constraint on the number of parallel edges between each pair of linked vertices in the $k$-core, namely, they should have at least $h$ parallel edges. Li et al~\cite{li2018persistent} proposed the persistent community search problem and gives a complicated instance called $(\theta,\tau)$-persistent $k$-core, which is a $k$-core persists over a time interval whose span is decided by the parameters. Similarly, Li et al~\cite{li2021efficient} proposed the continual cohesive subgraph search problem. Chu et al~\cite{chu2019online} studied the problem of finding the subgraphs whose density accumulates at the fastest speed, namely, the subgraphs with bursting density. Qin et al~\cite{qin2020periodic, qin2019mining} proposed the periodic community problem to reveal frequently happening patterns of social interactions, such as periodic $k$-core. Wen et al~\cite{bai2020efficient} relaxed the constraints of $(k,h)$-core and proposed quasi-$(k,h)$-core for better support of maintenance. Lastly, Ma et al~\cite{ma2019efficient} studied the problem of finding dense subgraph on weighted temporal graph. These works all focus on some specific patterns of cohesive substructure on temporal graphs, and propose sophisticated models and methods. Compared with them, our work addresses a fundamental querying problem, which finds the most general $k$-cores on temporal graphs with respect to two basic conditions, namely, $k$ and time interval. As discussed in Section~\ref{sec:extquery}, we can extend TCQ to find the more specific $k$-cores by importing the constraints defined by them, because most of the definitions are special cases of temporal $k$-core, but not vice versa.

Lastly, many research work on cohesive subgraph query for non-temporal graphs also inspire our work. We categorize them by the types of graphs as follows: undirected graph~\cite{huang2014querying, bonchi2019distance, liu2021local, zhang2020exploring, fang2020survey, yao2021efficient}, directed graph~\cite{sozio2010community, ma2020efficient, chen2021efficient}, labeled graph~\cite{sun2020stable, li2015influential, dong2021butterfly}, attributed graph~\cite{islam2022keyword, huang2017attribute, matsugu2019flexible, fang2017effectiveattr}, spatial graph~\cite{fang2017effectivespat, fang2018spatial, zhu2017geo}, heterageneous information network~\cite{fang2020effective}. Besides, many work specific to bipartite graph~\cite{wang2021efficient, zhang2021pareto, wang2021discovering, liu2019efficient} also contain valuable insights.

\section{Conclusion and Future Work}

For querying communities like $k$-cores on temporal graphs, specifying a time interval in which the communities emerge is the most fundamental query condition. To the best knowledge we have, we are the first to study a temporal $k$-core query that allows the users to give a flexible interval and returns all distinct $k$-cores emerging in any subintervals. Dealing with such a query in brute force is obviously costly due to the possibly large number of subintervals. Thus, we propose a novel decremental $k$-core inducing algorithm and the auxiliary optimization and implementation methods. Our algorithm only enumerates the necessary subintervals that can induce a final result and reduces redundant computation between subintervals significantly. Moreover, the algorithm is physically decomposed to a series of efficient data structure manipulations. As a result, although our algorithm does not use any precomputed index, it still outperforms an incremental version of the latest index-based approach by a remarkable margin. In conclusion, our algorithm is scalable with respect to the span of given time interval.

In the future, we will study how to leverage our algorithm as a framework to integrate various temporal $k$-core analytics. There are a number of related work have considered different temporal constraints of $k$-cores, most of which can be combined with the time interval condition to offer more powerful functionality. However, their query processing algorithms are essentially diverse. Therefore, we need to bridge the gap based on a general and scalable algorithm like ours.

	\begin{comment}
	\begin{acks}
		This work was supported by the [...] Research Fund of [...] (Number [...]). Additional funding was provided by [...] and [...]. We also thank [...] for contributing [...].
	\end{acks}   
	\end{comment}

	%\clearpage
	
	\bibliographystyle{ACM-Reference-Format}
	\bibliography{sample.bib}
	
\end{document}